\documentclass[pdflatex,sn-mathphys-num]{sn-jnl}

\usepackage{graphicx}%
\usepackage{multirow}%
\usepackage{amsmath,amssymb,amsfonts}%
\usepackage{amsthm}%
\usepackage{mathrsfs}%
\usepackage[title]{appendix}%
\usepackage{xcolor}%
\usepackage{textcomp}%
\usepackage{manyfoot}%
\usepackage{booktabs}%
\usepackage{algorithm}%
\usepackage{algorithmicx}%
\usepackage{algpseudocode}%
\usepackage{listings}%
\usepackage{subfig}%

\theoremstyle{thmstyleone}%
\newcommand{\lidx}[2]{{\vphantom{#2}}^{#1}\!{#2}}

\theoremstyle{thmstyletwo}%

\theoremstyle{thmstylethree}%

\begin{document}

\title[Gravitational entropy of fluids with energy flux]{Gravitational entropy of fluids with energy flux}


\author*[1]{\fnm{Roberto A.} \sur{Sussman}}\email{sussman@nucleares.unam.mx}

\author*[2]{\fnm{Sebasti\'an} \sur{N\'ajera}}\email{sebastian.najera@icf.unam.mx}

\author[1]{\fnm{Fernando A.} \sur{Piza\~na}}\email{klesto92@ciencias.unam.mx}

\author[2]{\fnm{Juan Carlos} \sur{Hidalgo}}\email{hidalgo@icf.unam.mx}

\affil[1]{\orgdiv{Instituto de Ciencias Nucleares}, \orgname{Universidad Nacional Aut\'onoma de M\'exico
(ICN-UNAM)}, \orgaddress{\street{A. P. 70 –- 543}, \city{Ciudad de M\'exico}, \postcode{04510}, \state{}, \country{M\'exico}}}

\affil[2]{\orgdiv{Instituto de Ciencias F\'isicas}, \orgname{Universidad Nacional Aut\'onoma de M\'exico
(ICF-UNAM)}, \orgaddress{\street{}, \city{Cuernavaca}, \postcode{62210}, \state{Morelos}, \country{M\'exico}}}


\abstract{We examine gravitational entropy growth within the formalism of Clifton, Ellis and Tavakol (CET) applied to a class of spherically symmetric exact solutions whose source is a shear-free fluid with energy flux in a comoving frame. By considering these solutions as potential cosmological models, we update previous literature that considered them only as restricted toy models of radiating spheres collapsing in a Vaidya background. In the present paper we examine  the integrability of the CET entropy form in connection with Einstein's equations in the fluid flow approach, proving as well that  all expanding configurations  comply with the growth of CET gravitational entropy.  Finally, we examine the connection between the CET gravitational entropy and the notion of a gravitational ``arrow of time'' based on the ratio of Weyl to Ricci curvature. Some of the solutions also provide potentially useful and viable inhomogeneous generalizations of FLRW models, thus suggesting an appealing potential for applications to current cosmological research.}




\maketitle

\section{Introduction}\label{intro}

Observational evidence and theoretical considerations (the near homogeneity and isotropy of the CMB) \cite{aghanim2020planck, arbey2021dark, grohs2023big} strongly suggest the early universe up to matter-radiation decoupling to be close to thermal equilibrium. However, an entropy defined in terms of thermal interactions necessarily should reach a maximum and stop growing when structure formation sets in and the Universe becomes gravity dominated and non-thermal. 

Given this outlook, Penrose \cite{penrose1979singularities} conjectured that some sort of gravitational entropy notion was needed to account for structure formation that takes place when thermal processes that increased entropy in the early Universe plasma are no longer dominant. Penrose also remarked that the Weyl tensor can be thought of as encoding the free gravitational field, as it is nonzero even in the absence of sources, in which case the Ricci tensor vanishes (hence it is the curvature associated with the sources). 

Bearing in mind these theoretical connections, Penrose proposed the concept of a gravitational ``arrow of time'' that guides self-gravitating systems undergoing structure formation to evolve through timelike directions in which an early dominant Ricci curvature should be overtaken by an increasing Weyl curvature as the Universe evolved away from the early plasma into a non-linear structure formation process associated with the long range gravitational interaction. 

Penrose and other authors have speculated that when this ratio is non-decreasing it should signal a direction of increasing inhomogeneity, producing an increase of entropy in the sense of the application to self-gravitating systems in the framework of the microcanonical ensemble. As the system becomes more inhomogeneous a wider range of densities and momenta are accessed, which increase the volume of accessible states in the phase space. In more recent appraisals of Penrose's proposal \cite{pelavas2006gravitational, zhao2018black, guha2020gravitational, gregoris2022understanding} the Ricci scalars can be replaced by the Kretschmann scalar that is nonzero even in vacuum solutions (allowing for the application to vacuum spacetimes). 

The ``arrow of time'' proposal holds for known Petrov type D dust solutions along the comoving frame, with an increasing ratio of Weyl to Ricci scalar curvature as inhomogeneity increases and becomes completely encoded in the Weyl tensor. In fact, in inhomogeneous LTB and Szekeres models (for which the magnetic Weyl tensor vanishes) the components of the electric Weyl tensor are determined by the exact density deviation from a homogeneous density background. It is possible to prove that this relation between inhomogeneity and dominance of Weyl curvature also holds in LTB and Szekeres models with nonzero pressure. 

However, the ratio of Weyl to Ricci curvature might not increase as inhomogeneity grows in a comoving frame in spacetimes with more elaborate sources. In 1985 Bonnor \cite{bonnor1985gravitational} defined a ``thermodynamical arrow of time'' as a future direction of inhomogeneity growth, to distinguish with Penrose's ``gravitational arrow of time'' associated with the future direction of increasing ratio of Weyl to Ricci curvature. Bonnor proved that both ``arrows of time'' coincided for dust models, but not for heat conducting shear-free models that have been used to model collapsing radiating spheres in a Vaidya exterior background. For these solutions the ratio of Weyl to Ricci scalar curvatures decreases as inhomogeneity grows while collapse proceeds with increasing proper time. 

Rudjord et al \cite{ystein2008weyl} and more recently Chakraborty et al \cite{chakraborty2021investigation, chakraborty2024arrow} revisited and validated Bonnor's result for these models, considering also the ratio of the scalar $C_{abcd}C^{abcd}$ to the Kretschmann scalar to allow applying the ratio to vacuum spacetimes. Nevertheless, Bonnor's result was tested only on a class of shear-free fluids with energy flux described as heat conduction only to model radiating spheres collapsing in a Vaidya background. In the  the present article we show that Bonnor's result does not hold in the same class of solutions when considering them as expanding cosmological models not restricted by any exterior matching.

Further theoretical development on the role of Weyl curvature in structure formation was achieved by Clifton, Ellis and Tavakol (CET), whose gravitational entropy proposal \cite{clifton2013gravitational} also regards Weyl curvature as encoding the free gravitational field, but instead of proposing a scalar formulation they propose an analogue of the Gibbs one-form constructed from an effective gravitational energy-momentum tensor that emerges from the algebraic decomposition of the Bel-Robinson tensor, which is the only divergence-free tensor that can be constructed from the Weyl tensor. The CET entropy proposal is  theoretically more robust than previous formulation by Penrose and other in terms of curvature scalars.

However, the Bel-Robinson tensor is fourth order, hence CET considered as a candidate for the effective energy-momentum tensor the second order tensors that result from the algebraic decomposition of the Bel-Robinson tensor as a tensor product of two second order tensors (i.e. the ``square root'' of the Bel Robinson tensor). This algebraic decomposition had been examined by Bonilla and Senovilla \cite{bonilla1997some, bonilla1997very, bonilla1998miscellaneous}, showing that it is only unique and non-degenerate for Petrov types N and D (wavelike and Coulomb-like spacetimes), with a degeneracy of non-unique second order tensors for the remaining Petrov types. 

The unique second order root of the Bel-Robinson tensor for spacetimes of Petrov types D and N, can be invariantly projected in terms of an arbitrary 4-velocity field and its orthogonal rest spaces, leading to the definition of "gravitational" state variables: density, energy flux and stress tensor, with these variables unrelated to the state variables from the energy-momentum tensor of the sources \cite{pelavas2006gravitational, pelavas2000measures}. From these geometric state variables, CET construct the 4-velocity projected component of an analogue of the thermodynamical Gibbs one form, that defines a convective derivative of a gravitational entropy, with a gravitational temperature as its integrating factor. Since this procedure does not determine the gravitational temperature (as would be the case with an equation of state in a thermal system), CET introduce an ad hoc gravitational temperature in terms of the definition of gravitational redshift between local comoving observers. 

Once the geometric state variables have been determined for a given spacetime (i.e. a solution of Einstein's equations), the next task in the CET proposal is to verify if the gravitational entropy increases with proper time. This task has been accomplished for LTB \cite{sussman2015gravitational} and Szekeres class I dust models \cite{pizana2022gravitational}, showing that CET entropy grows in inhomogeneities (over-densities and voids) that can be described by the exact generalization of the growing mode of dust perturbations. In particular, \cite{sussman2015gravitational, pizana2022gravitational} show that a positive cosmological constant provides a finite asymptotic saturation value for the CET gravitational entropy. However, other references that have examined the CET entropy  \cite{gregoris2020thermodynamics} show that the magnitude of the ratio of Weyl to Ricci curvature decreases with cosmic expansion, a result that contradicts the assumptions of the CET proposal. Furthermore, in \cite{chakraborty2022appropriate} CET entropy has also been applied to wormholes and compared to other gravitational entropy definitions, providing a unique gravitational entropy for Petrov D and N spacetimes. 

The extensive literature \cite{de1985collapse,santos1985non, kolassis1988energy, bonnor1989radiating,grammenos1995radiating} looking at collapsing spheres matched to a Vaidya exterior relied on extremely simplified metric ansatzes to model these spheres as sheer-free heat conducting  fluids, attempting to interpret the heat flux in terms of constitutive equations of non-equilibrium thermodynamics   \cite{maharaj2012radiating, schafer2000gravitational}. This literature see a summary in  \cite{tewari2015dissipative} includes the toy models used by Bonnor and other authors \cite{bonnor1985gravitational, chakraborty2024arrow} to test Penrose's proposal.  

In this article we abandon the thermodynamical approach of previous literature and examine these solutions as potential cosmological models that are based on more general extra free parameters and solutions instead of  simplified ad hoc ansatzes. We also  examine the integrability of the Gibbs one-form in connection with Einstein's equation in the 1+3 fluid flow approach, showing that the gravitational variables of CET can be related to the 4-acceleration and energy flux that are the variables driving the inhomogeneity of the models.  We show that CET entropy grows in all expanding configurations based on specific particular solutions that fully comply with regularity and asymptotic conditions. This result explains the contradiction between CET entropy growth and Penrose's "arrow of time'' (noted by Bonnor and others): this contradiction only holds in the context of a collapse in a Vaidya exterior, but does not occur in the cosmological context we have considered. 

The section by section content of the paper is described as follows: In Sec.~\ref{sec:shearfree}, we introduce and summarize shear-free spherically symmetric models with variables that facilitate their interpretation in a cosmological scenario. In Sec.~\ref{sec:CET} we outline the gravitational entropy proposal by Clifton, Ellis and Tavakol (CET). Sec.~\ref{CETGR} examines the relation between the CET equations and Einstein's equations within the ``1+3'' fluid flow formulation , while Sec.~\ref{sec:testingCET} applies the CET formalism to the shear-free solutions developed earlier, showing that these models can satisfy a gravitational entropy growth as they expand in time, In Sec.~\ref{sec:timeevol}, we discuss the time evolution of these models by setting an FLRW-like scale factor as a free parameter. In Sec.~\ref{sec:arrow}, we analyze Penrose's ``arrow of time'' criterion and show that the ratio of Weyl to Ricci curvature scalars is increasing with time in some cosmological scenarios. In Sec.~\ref{sec:conclusions}, we discuss the obtained results and suggest future applications and extensions. Finally, we provide four appendices: In Appendix \ref{sec:shearfreeapp} provides details on the derivation of the shear-free solutions, Appendix \ref{sec:nonconflat} presents in detail the metrics and properties of the specific solutions used in Sec.~\ref{sec:testingCET}. Appendix \ref{regularity} discusses regularity conditions and Appendix \ref{sec:1p3eq} summarizes the ``1+3'' evolution and constraint equations for the solutions under consideration.

\section{Shear-free models with energy flux}\label{sec:shearfree}
We consider spherically symmetric shear-free solutions whose energy-momentum tensor has nonzero energy flux in a comoving frame. This energy flux has been described in the literature as heat conduction in a thermodynamical framework for models of radiating spheres in a Vaidya background. Since our aim is to consider these solutions as potentially viable cosmological models, we describe them in a different coordinate representation that is more suitable for a cosmological context (see additional details Appendix \ref{sec:shearfreeapp} ) 
 \begin{eqnarray}
 ds^2&=&\frac{-N^2 dt^2+d\chi^2+ f^2(\chi)(d\theta^2+\sin^2\theta d\phi^2)}{L^2},\label{metric}\\
 f(\chi) &=& \left\{  \begin{array}{ll}
\chi & k_0=0 \\
          \sin \chi& k_0=1 \\
         \sinh \chi & k_0=-1 
\end{array}    \right.,\label{eqf}\\
 T_{ab}&=&  \rho u_a u_b + p h_{ab}+2q_{(a}u_{b)},\label{Tab}    
 \end{eqnarray} 
 where $h_{ab}=g_{ab}+u_au_b$,\,\,$q_au^a=0$ and $N,\,L$ and $\rho,\,p$, as well as the components of the energy flux $q_a$, depend on $(t, \chi)$, while $k_0=\epsilon_0H_0^2$ with $\epsilon_0=0,\pm 1$ (we will write henceforth $k_0=0,\pm 1$)\footnote{The constant $k_0$ lacks the geometric interpretation of "curvature index" in FLRW models, as it is not necessarily proportional to the curvature scalar $\lidx{3}{R}$ of the constant time spatial hypersurfaces. }. The radial coordinate $\chi$ has the following ranges: $0\leq \chi<\infty$ for $k_0=0,-1$ but $0\leq \chi\leq \pi$ for $k_0=1$. All previously published heat conducting models considered only the case $k_0=0$, but the cases $k_0=\pm 1$ in (\ref{eqf}) incorporate useful degrees of freedom in a cosmological approach.
 
The isotropic pressure condition $G^r_r-G^\theta_\theta=0$ (see Eq.~\eqref{CPI1}) implies the following coupled linear system on $N$ and  $L$
\begin{equation} N_{,yy}- J\,N=0,\qquad  2L_{,yy}-J\,L=0,\label{pdsNL} \end{equation}
where the variable $y$ is defined as
\begin{eqnarray}  
f^2(\chi) &=& y(2-k_0 y)\quad\Rightarrow\quad 
      y(\chi)= \left\{  \begin{array}{ll} 
         \chi^2/2,& k_0=0,  \\
         1-\cos\,\chi,& k_0=1, \\
         \cosh\,\chi-1, & k_0=-1.  
       \end{array}\right.\label{ychi}
\end{eqnarray}       
and $J=J(y)$ is an arbitrary free function\footnote{It is possible to assume $J=J(t,y)$, but for the sake of simplicity we will not assume this time dependence.} acting as a generating function to derive exact solutions.

Being solutions of a linear system, each one of $L$ and $N$ in the metric Eq.~\eqref{metric} are always expressible as the a linear combination of two functions of $y$, with 4 functions of time appearing as integration constants, leading to 
\begin{equation}\frac{-[\nu_1(t) N_1(y)+\nu_2(t) N_2(y)]^2 dt^2+d\chi^2+ f^2(\chi)(d\theta^2+\sin^2\theta d\phi^2)}{[\lambda_1(t)L_1(y)+\lambda_2(t) L_2(y)]^2}.\label{LN} \end{equation}
However, we simplify further ahead this metric by coordinate rescalings and a redefinition of the time dependent functions.     

Since the case $J=0$ leads to conformally flat solutions with zero Weyl tensor, we henceforth assume that $J\ne 0$. The simplest choice of non-conformally flat solutions follows from Eqs.~(\ref{pdsNL}), (\ref{LN}) and (\ref{metric}), with $J=\epsilon_0\Delta^2$ where $\Delta>0$ is an arbitrary positive constant
\begin{equation} N_{,yy}- \epsilon_0 \Delta^2\,N=0,\qquad L_{,yy}-\frac{\epsilon_0}{2}\Delta^2\,L=0,\qquad \epsilon_0=0,\pm1 ,\label{pdsNL2} \end{equation}
admitting for $\epsilon_0=\pm 1$ the following exact solutions for the generic metric (\ref{metric})
\begin{eqnarray}
 \epsilon_0=1\qquad   \left\{ \begin{array}{ll}
    N= \nu_1(t)\,\cosh(\Delta\,y)+\nu_2(t)\, \sinh(\Delta\,y), \\
    L= \lambda_1(t)\,\cosh(\Delta\,y/\sqrt{2})+\lambda_2(t)\, \sinh(\Delta\,y/\sqrt{2})
       \end{array}\right.,  \label{epos}   
 \end{eqnarray}
\begin{eqnarray}
\epsilon_0=-1\qquad \left\{ \begin{array}{ll}
         N= \nu_1(t)\,\cos(\Delta\,y) + \nu_2(t)\,\sin(\Delta\,y), \\
         L= \lambda_1(t)\,\cos(\Delta\,y/\sqrt{2}) + \lambda_2(t)\,\sin(\Delta\,y/\sqrt{2}),  
       \end{array}\right.,\label{eneg}
       \end{eqnarray}
 where $y=y(\chi)$ is given by (\ref{ychi}). There are 6 different solutions contained in (\ref{epos})-(\ref{eneg}): for each sign of $ \epsilon_0=\pm 1$  there are 3 cases for $k_0=0,\pm 1$ in (\ref{ychi}). A detailed analysis of the cases (\ref{epos}) and (\ref{eneg}) can be found in Appendix \ref{sec:shearfree}. 
 
To apply the solutions Eq.~\eqref{LN} and Eqs.~\eqref{epos}-\eqref{eneg} to  a cosmological context, it is convenient to transform the metric Eq.\eqref{LN} into
 \begin{equation}ds^2=a^2(\eta)\left[\frac{-[N_1(y)+\nu(\eta) N_2(y)]^2 d\eta^2+d\chi^2+ f^2(\chi)(d\theta^2+\sin^2\theta d\phi^2)}{[L_1(y)+\lambda(\eta) L_2(y)]^2}\right].\label{LN2} \end{equation}
 where $d\eta=\nu_1 dt$ and $a=1/\lambda_1$, while $\nu=\nu_2/\nu_1$ and $\lambda=\lambda_2/\lambda_1$. Unless specified otherwise, we use this parametrization of the metric, though the choice of $\nu_1,\,\lambda_1$ to define $\eta$ and $a$ is arbitrary (it is possible to do these re-scalings with either one of the functions in $L$ and $N$, see  Appendix \ref{sec:shearfreeapp}).
 
 \section{The gravitational entropy proposal of Clifton, Ellis and Tavakol (CET)}\label{sec:CET}

CET consider the Bel-Robinson tensor, the unique divergence-less tensor that can be constructed from the Weyl tensor:
\begin{equation} {\cal T}_{abcd}=\frac14\left(C_{eabf}C^{ef}_{cd}+C^*_{eabf}C^{*ef}_{cd}\right),\label{BelRob}\end{equation}
where $C^*_{abcd}=\frac12 \eta_{abef}C^{ef}_{cd}$ is the dual Weyl tensor. A second order symmetric divergence-free tensor $t_{ab}$ can be obtained from a fourth order symmetric divergence-free tensor in general ${\cal F}_{abcd}$, from the algebraic decomposition known as its ``square root'' \cite{bonilla1997some, bonilla1997very, bonilla1998miscellaneous}
\begin{eqnarray} {\cal F}_{abcd}=t_{(ab}t_{cd)}-\frac12 t_{e(a}t_b^eg_{cd)}-\frac14 t_e^et_{(ab} g_{cd)}+\frac{1}{24}\left(t_{ef}t^{ef}+\frac12 (t_e^e)^2\right)g_{(ab}g_{c)}.\label{decomp}\end{eqnarray}
While every symmetric divergence-free $t_{ab}$ leads to a unique fourth order symmetric divergence-free tensor ${\cal F}_{abcd}$ through (\ref{decomp}), the converse statement is false. Given the fourth order tensor ${\cal F}_{abcd}$, there is no (in general) unique $t_{ab}$. In particular, for the Bel-Robinson tensor (\ref{BelRob}), a unique second order symmetric divergence-free tensor is only possible for spacetimes of Petrov type D and N. Since the models we are considering are Petrov type D, we will use the ``square root'' derived by CET for these spacetimes in an orthonormal tetrad $\{x_{_\textrm{\tiny{A}}},\,y_{_\textrm{\tiny{A}}},\,z_{_\textrm{\tiny{A}}},\,u_{_\textrm{\tiny{A}}}\}$ with $\textrm{\tiny{A}}=0,1,2,3$:
\begin{equation}8\pi {\cal T}_{_{_\textrm{\tiny{AB}}}}=\epsilon\alpha |\Psi_2|\left[x_{_\textrm{\tiny{A}}}x_{_\textrm{\tiny{B}}} +y_{_\textrm{\tiny{A}}}y_{_\textrm{\tiny{B}}} -2(z_{_\textrm{\tiny{A}}}z_{_\textrm{\tiny{B}}} -u_{_\textrm{\tiny{A}}} u_{_\textrm{\tiny{A}}})\right],\label{TT} \end{equation}
where $\epsilon=\pm 1$,\,\, $\alpha$ is a constant (to set units) and $\Psi_2= C_{_{_\textrm{\tiny{ABCD}}}}\,k^{_\textrm{\tiny{A}}}\,m^{_\textrm{\tiny{B}}}\,\tilde m^{_\textrm{\tiny{C}}}\,l^{_\textrm{\tiny{D}}}$ is the only nonzero Weyl scalar for Petrov type D spacetimes given in terms of the null tetrad associated with $\{x_{_\textrm{\tiny{A}}},\,y_{_\textrm{\tiny{A}}},\,z_{_\textrm{\tiny{A}}},\,u_{_\textrm{\tiny{A}}}\}$.

Following CET and considering that all energy-momentum tensors are second order symmetric and divergence-less, we consider ${\cal T}_{_{_\textrm{\tiny{AB}}}}$ in (\ref{TT}) as an ``effective'' energy-momentum tensor associated with Weyl curvature ({\it i.e.} free gravitational field). It is important to remark that ${\cal T}_{_{_\textrm{\tiny{AB}}}}$ does not represent an energy-momentum tensor of a source to be placed in the right hand side of Einstein's equations, but a formal geometric energy-momentum tensor. Nevertheless, given a 4-velocity field $u^{_\textrm{\tiny{A}}}$ and its orthogonal rest space projection $h^{_{_\textrm{\tiny{AB}}}}=u^{_\textrm{\tiny{A}}}\,u^{_\textrm{\tiny{B}}}+\eta^{_{_\textrm{\tiny{AB}}}}$, this energy-momentum tensor can define ``gravitational state variables'' ({\it i.e.} gravitational density, pressure, anisotropic pressure and energy flux) .
\begin{equation}{\cal T}^{{\textrm{\tiny{AB}}}}=\rho_{_\textrm{\tiny{gr}}}u^{_\textrm{\tiny{A}}}\,u^{_\textrm{\tiny{B}}}+p_{_\textrm{\tiny{gr}}}h^{_{_\textrm{\tiny{AB}}}}+\Pi_{_\textrm{\tiny{gr}}}^{{_\textrm{\tiny{AB}}}}+2 q_{_\textrm{\tiny{gr}}}^{_\textrm{\tiny{(A}}}u^{_\textrm{\tiny{B)}}},\label{Tabgr}\end{equation}
where $\Pi_{_\textrm{\tiny{gr}}}^{{\textrm{\tiny{AB}}}}=\left[h^{{\textrm{\tiny{A}}}}_{_\textrm{\tiny{C}}}h^{{\textrm{\tiny{B}}}}_{_\textrm{\tiny{D}}}-\frac13 h^{{\textrm{\tiny{AB}}}}h_{_\textrm{\tiny{CD}}}\right]{\cal T}^{{\textrm{\tiny{CD}}}}$. For the specific tensor (\ref{TT}) and setting up units, we have
\begin{eqnarray}16\pi \rho_{_\textrm{\tiny{gr}}}&=&|\Psi_2|,\quad 16\pi \Pi^{\textrm{\tiny{gr}}}_{_{\textrm{\tiny{AB}}}}= |\Psi_2|\left[-x_{_\textrm{\tiny{A}}}x_{_\textrm{\tiny{B}}} +y_{_\textrm{\tiny{A}}}y_{_\textrm{\tiny{B}}} +z_{_\textrm{\tiny{A}}}z_{_\textrm{\tiny{B}}} +u_{_\textrm{\tiny{A}}} u_{_\textrm{\tiny{A}}}\right],\nonumber\\  p_{_\textrm{\tiny{gr}}}&=&q_{_\textrm{\tiny{gr}}}^{_\textrm{\tiny{A}}}=0,\label{TTCET}
\end{eqnarray}  
Since the gravitational energy density $\rho_{_\textrm{\tiny{gr}}}$ is associated with Weyl curvature (the ``free gravitational field''), its corresponding energy ${\cal E}_{_\textrm{\tiny{gr}}}$ (in geometric units) can be defined by analogy with the local energy in equilibrium thermal systems as the product of the gravitational energy density multiplied by the local proper spatial volume ($V$ such that $\dot V/V=\Theta=\nabla_uu^a$)
\begin{equation}  {\cal E}_{_\textrm{\tiny{gr}}} =   16\pi\rho_{_\textrm{\tiny{gr}}}\,V=|\Psi_2|\,V,\label{Egrav}\end{equation} 
where $\rho_{_\textrm{\tiny{gr}}}$ is given by (\ref{TTCET}). By analogy with the Gibbs one form $T{\bf d} S = {\bf d} {\cal E}+ p {\bf d} V$, where ${\cal E}=\rho V$ is the total energy in a volume $V$ and $T$ is the temperature, we bear in mind that $p_{_\textrm{\tiny{gr}}} =0$ and project the Gibbs form of the gravitational variables on the 4-velocity field in (\ref{TTCET}): 
\begin{equation} T_{_\textrm{\tiny{gr}}}\dot S_{_\textrm{\tiny{gr}}}=T_{_\textrm{\tiny{gr}}} u^a \nabla_a S_{_\textrm{\tiny{gr}}} =  \dot {\cal E}_{_\textrm{\tiny{gr}}}=u^a \nabla_a( E_{_\textrm{\tiny{gr}}}),\label{Gibbs}\end{equation}
The analogy with the Gibbs one-form provides no information on $T_{_\textrm{\tiny{gr}}}$, thus CET define it in terms of local redshift for comoving observers expressed in terms of the kinematic parameters that follow from the decomposition of $\nabla_b u_a$
\begin{equation}  T_{_\textrm{\tiny{gr}}} =\left| \nabla_{_\textrm{\tiny{A}}} u_{_\textrm{\tiny{B}}} k^{_\textrm{\tiny{A}}}\,l^{_\textrm{\tiny{A}}}\right|,\label{Tgrav}\end{equation}
where $k^{_\textrm{\tiny{A}}}\,\,l^{_\textrm{\tiny{A}}}$ are the null vectors associated with the tetrad $\{x_{_\textrm{\tiny{A}}},\,y_{_\textrm{\tiny{A}}},\,z_{_\textrm{\tiny{A}}},\,u_{_\textrm{\tiny{A}}}\}$ and $ \nabla_{_\textrm{\tiny{A}}} u_{_\textrm{\tiny{B}}}$ becomes in a coordinate basis
 \begin{equation} \nabla_a u_b = \frac13\Theta h_{ab}-\dot u_a u_b +\sigma_{ab}+\omega_{ab},\label{GradU}\end{equation}
identifying the expansion scalar $\Theta=\nabla_a u^a$, the 4-acceleration $\dot u_a = u^b \nabla_b u_a$, the shear $\sigma_{ab}=\nabla_{(a} u_{b)}-\frac13 \Theta h_{ab}$ and $\omega_{ab}=\nabla_{[a} u_{b]}$ vorticity tensors. 

Since $T_{_\textrm{\tiny{gr}}}\geq 0$ by construction, the necessary and sufficient condition for gravitational entropy production becomes 
\begin{equation}   \dot {\cal E}_{_\textrm{\tiny{gr}}}= u^a \nabla {\cal E}_{_\textrm{\tiny{gr}}}\geq 0,\label{entprod0}\end{equation}
but to evaluate the gravitational entropy $S_{_\textrm{\tiny{gr}}}$ we need to integrate the Gibbs form along $u^a$: 
\begin{equation}   \dot S_{_\textrm{\tiny{gr}}}=\frac{\dot {\cal E}_{_\textrm{\tiny{gr}}}}{T_{_\textrm{\tiny{gr}}}}= \frac{u^a \nabla {\cal E}_{_\textrm{\tiny{gr}}}}{T_{_\textrm{\tiny{gr}}}}\label{entprod2}.\end{equation}
In what follows we explore the connections between (\ref{entprod0})-(\ref{entprod2}) and Einstein's equations for the models under consideration.

\section{The CET gravitational entropy and Einstein's equation}\label{CETGR}

It is evident that CET gravitational entropy  is closely linked to the dynamics of the Weyl tensor, which for Petrov type D models under examination reduce to its electric part and specifically to the conformal Weyl invariant $\Psi_2$, which for the metric \eqref{metric} takes the form
\begin{equation} E^a_b= \Psi_2\,{\bf e}^a_b,\qquad \Psi_2 = -\frac16 (f^2\,L^2\,J),\label{Edef}\end{equation}
where ${\bf e}^a_b$ is the covariantly constant tensorial base for Petrov type D spacelike symmetric traceless tensors. In the coordinates of \eqref{metric} it is ${\bf e}^a_b=h^a_b-3n^an_b=\hbox{diag}[0,-2,1,1]$ for $n_an^a=1$ and $n_au^b=0$. This tensor complies with the following properties: ${\bf e}^a_a = g_{ab}{\bf e}^{ab}=0,\,\,\, {\bf e}^{ab}{\bf e}_{ab}=6$ and  $\nabla_c{\bf e}^a_b=0$, plus 
\begin{eqnarray} 
{\bf e}_{\langle ab\rangle}=\left[h_a^c h_b^d-\frac13 h_{ab}h^{cd}\right]{\bf e}_{cd}={\bf e}_{ab},\quad A_{\langle ab\rangle}{\bf e}^{ab}=A_{ab}{\bf e}^{ab},\label{props}\end{eqnarray}
In particular, Eqs.~\eqref{Egrav}, \eqref{Gibbs}, \eqref{entprod} and \eqref{entprod2} denote a close relation with the time  evolution of the electric Weyl tensor. Therefore, these equations must be connected to the time evolution law for $E^a_b$ furnished by General Relativity for the irrotational shear-free models under consideration, restricted by $\sigma_{ab}=\pi_{ab}=H_{ab}=\omega_{ab}=0$. The  evolution law under the fluid flow 1+3 formalism is 
\begin{equation}\dot{E}_{\langle ab\rangle} + \Theta E_{ab} = \dot E_{cd}{\bf e}_a^c{\bf e}_b^d+\Theta E_{ab}=-4\pi ( \nabla_{\langle a}q_{b\rangle} 
   +2 \dot{u}_{\langle a}q_{b\rangle}),\label{dotE}\end{equation}
The connection between this evolution law and the CET entropy equations \eqref{entprod} and \eqref{entprod2} follows from the contraction
 \begin{equation} 6\Psi_2=E_{ab}{\bf e}^{ab}\end{equation}
which  provides through \eqref{TTCET} a direct link between the gravitational energy density (in geometric units)  and the electric  Weyl tensor in \eqref{Edef} 
\begin{equation}16\pi \rho_{_\textrm{\tiny{gr}}}=|\Psi_2| = \frac16 |E_{ab}{\bf e}^{ab}|\quad\Rightarrow\quad {\cal E}_{_\textrm{\tiny{gr}}}=16\pi \rho_{_\textrm{\tiny{gr}}}\,V=|\Psi_2| V=\frac16 |E_{ab}{\bf e}^{ab}|\,V.\label{EEWeyl}\end{equation} 
From \eqref{Edef} it is evident that the sign of $\Psi_2$ depends on the sign of the generatrix function $J(y)$  (which we assume nonzero) that determines the solutions. To handle the absolute value in ${\cal E}_{_\textrm{\tiny{gr}}}$ we assume two cases: $J>0$ and $J<0$, bearing in mind that $\Theta=\dot V/V$ and that ${\bf e}^{ab}$ is covariantly constant ($\nabla_c{\bf e}^{ab}=0$)
\begin{itemize}
\item $J<0,\,\,\Psi_2>0$, hence ${\cal E}_{_\textrm{\tiny{gr}}}= \frac16 E_{ab}{\bf e}^{ab}\,V,$:
\begin{eqnarray} \dot {\cal E}_{_\textrm{\tiny{gr}}}=\frac16 (\dot E_{ab} V+E_{ab}\dot V){\bf e}^{ab}=\frac16(\dot E_{ab}+\Theta E_{ab})V{\bf e}^{ab}=-\frac{2\pi}{3}( \nabla_{a}q_{b} 
   +2 \dot{u}_{ a}q_{b}){\bf e}^{ab},\label{dotEneg}
\end{eqnarray}
\item $J>0,\,\,\Psi_2>0$, hence ${\cal E}_{_\textrm{\tiny{gr}}}= -\frac16 E_{ab}{\bf e}^{ab}\,V,$:
\begin{eqnarray} \dot {\cal E}_{_\textrm{\tiny{gr}}}=-\frac16 (\dot E_{ab} V+E_{ab}\dot V){\bf e}^{ab}=-\frac16(\dot E_{ab}+\Theta E_{ab})V{\bf e}^{ab}=\frac{2\pi}{3}( \nabla_{a}q_{b} 
   +2 \dot{u}_{ a}q_{b}){\bf e}^{ab},\label{dotEpos}
\end{eqnarray}
\end{itemize}
where we have used \eqref{props} to remove the angle brackets $\langle ab\rangle$ in \eqref{dotE}, \eqref{dotEneg} and \eqref{dotEpos}.  

Equations \eqref{dotEneg} and \eqref{dotEpos} provide an alternative form to express the gravitational entropy production law \eqref{entprod2} in relation with the 4-acceleration and energy flux ($\dot u_a,\,q_a$), the main physical effects behind the inhomogeneity of the models that should influence structure formation, We can now present the law of entropy production \eqref{entprod2} as
\begin{eqnarray}
\dot S_{_\textrm{\tiny{gr}}}=\frac{\dot {\cal E}_{_\textrm{\tiny{gr}}}}{T_{_\textrm{\tiny{gr}}}}=\pm \frac{2\pi}{3}\frac{( \nabla_{ a}q_{b} 
   +2 \dot{u}_{a}q_{b}){\bf e}^{ab}}{\left|\left(\frac13\Theta h_{ab} -\dot u_a u_b\right)k^a l^b\right|},\label{entprod3}
\end{eqnarray}
where the $\pm$ sign distinguishes the cases $J$ positive or negative. This form of the entropy production law relates  gravitational entropy expressed as a Gibbs one-form with a Weyl tensor related gravitational energy and gravitational  temperature with the main quantities behind the structure formation that should occur fro the inhomogeneity produced by the 4-acceleration and energy flux.  

However, besides the evolution equation \eqref{dotE}, Einstein's equations (see Appendix \ref{sec:1p3eq}) in  the 1+3 fluid flow approach contain the following two constraints  involving $E^a_b$, an algebraic one resulting  from the restriction $\sigma_{ab}=\pi_{ab}=H_{ab}=\omega_{ab}=0$ and the other involving a divergence of $E^a_b$
 \begin{eqnarray}
\nabla^b E_{ab} &=& \frac{4\pi}{3}(2\Theta q_a- 2\nabla_a \rho ),\label{divE}\\
E_{ab} &=& \nabla_{\langle a}\dot{u}_{b\rangle} 
  - \dot{u}_{\langle a}\dot{u}_{b\rangle}.\label{EWdef}  
\end{eqnarray}
Evidently, the algebraic definition  \eqref{EWdef} of $E_{ab}$ (which involves the scalar function $\Psi_2$) must be consistent with  the integrability conditions that follow from mixed derivatives in \eqref{dotE} and \eqref{divE}.  

For the metric \eqref{metric} with the metric functions satisfying \eqref{pdsNL} and the electric Weyl tensor given by \eqref{dotE}, the expansion scalar is $\Theta=3L_{,t}/N$ (see \eqref{Tht}), while the 4-acceleration and energy flux vectors are given by  \eqref{Accapp} and \eqref{eqQapp} in terms of the scalars ${\cal A}$ and $Q$. Therefore, for the metric \eqref{metric}, the 1+3 equations \eqref{dotE} and \eqref{divE}-\eqref{EWdef} become 
\begin{eqnarray}
\frac{L}{N} (\Psi_2)_{,t}+ \Theta \Psi_2&=&- \frac{4\pi}{3} \left(2 Q \mathcal{A}- \frac{Q f_{,\chi}}{ f} + Q_{,\chi} \right) L, \label{W_prop}\\
(\Psi_2+ \frac{\kappa}{6} \rho)_{,\chi} &=& \frac{4\pi}{3} \Theta Q + 3\Psi_2\frac{ f_{,\chi}}{f}, \label{W_constr}\\
\Psi_2 &=& -\frac{1}{3}\left(\mathcal{A}^2 + \frac{\mathcal{A}f_{,\chi}}{f} -  \mathcal{A}_{,\chi} \right)L^2.\label{EWdef2}
\end{eqnarray}
From these equations it is straightforward (but algebraically laborious) to show that $\Psi_2$ in either form \eqref{Edef} or \eqref{EWdef2} is consistent with the integrability condition  from the mixed derivatives that follow from \eqref{W_prop} and \eqref{W_constr}. This means that the evolution equation for the gravitational energy ${\cal E}_{_\textrm{\tiny{gr}}}$, is integrable either in the form  \eqref{entprod0} or \eqref{dotEneg}-\eqref{dotEpos}, an integrability that follows from Einstein's equations in the 1+3 formulation. However, the entropy one-form  \eqref{entprod2} in its form  \eqref{entprod3} involves the gravitational temperature $T_{_\textrm{\tiny{gr}}}$ in \eqref{Tgrav}, which bears no relation to Einstein's equations, hence probing its integrability requires further work that we leave for a future paper.

\section{Testing the CET entropy}\label{sec:testingCET} 

The gravitational state variables in equations (\ref{Tabgr})-(\ref{Gibbs}) are entirely determined by the gravitational entropy density $\rho_{_\textrm{\tiny{gr}}}$, equal to the conformal invariant $\Psi_2$ given by \eqref{Edef}. The gravitational entropy density $\rho_{_\textrm{\tiny{gr}}}$ in (\ref{TTCET}), local proper volume $V$ (such that $\dot V/V=\Theta=\nabla_a u^a$) and local gravitational entropy $S_{_\textrm{\tiny{gr}}}$ in (\ref{Egrav}) for the metric Eq.~\eqref{LN2} take the forms 
\begin{eqnarray} 16\pi \rho_{_\textrm{\tiny{gr}}}&=&|\Psi_2|=\frac{|\epsilon_0|  L^2 f^2\,\Delta^2}{6\,a^2},\qquad V= \frac{a^3\,f^2}{ L^3}\nonumber\\
&\Rightarrow&\quad {\cal E}_{_\textrm{\tiny{gr}}}= |\Psi_2|\,V=16\pi \rho_{_\textrm{\tiny{gr}}}\,V=\frac{|\epsilon_0| \Delta^2 a f^4}{6 L},\label{rhoVS}\end{eqnarray}
The null tetrad $k^A,\,l^A$ in the coordinate basis $(\eta,\,\chi,\,\theta,\,\varphi)$ of (\ref{LN2}) is
\begin{equation} k_A^a = x_A^a + y_A^a = \frac{ L}{aN}\delta^a_\eta+ \frac{L}{a}\delta^a_\chi,\qquad l_A^a = x_A^a - y_A^a = \frac{ L}{aN}\delta^a_\eta- \frac{L}{a}\delta^a_\chi,\end{equation}
together with $u_{a;b}=\frac13\Theta h_{ab}-\dot u_a u_b$ and (\ref{Tht})-(\ref{Accapp}) lead to
\begin{equation}T_{_\textrm{\tiny{gr}}}= \left| \nabla_{_\textrm{\tiny{A}}} u_{_\textrm{\tiny{B}}} k^{_\textrm{\tiny{A}}}\,l^{_\textrm{\tiny{B}}}\right|=\left| -\tilde L'+\frac{\tilde L N'}{N} +\frac{\tilde L_{,t}}{N}\right|.\label{Tgrav2}\end{equation}
where $\tilde L=L/a$ and   $\tilde L_{,t}=a\,\tilde L_{,\eta}$. Entropy production conditions (\ref{entprod0}) and (\ref{entprod2}) become
\begin{equation}    \dot {\cal E}_{_\textrm{\tiny{gr}}} =\frac{L}{aN}\,\frac{\partial {\cal E}_{_\textrm{\tiny{gr}}} }{\partial \eta}\geq 0,\qquad  \dot S_{_\textrm{\tiny{gr}}} = \frac{\dot {\cal E}_{_\textrm{\tiny{gr}}}}{T_{_\textrm{\tiny{gr}}}}=\frac{1}{T_{_\textrm{\tiny{gr}}}}\frac{L}{aN}\,\frac{\partial {\cal E}_{_\textrm{\tiny{gr}}}}{\partial \eta}\geq 0,\label{entprod}\end{equation}
where ${\cal E}_{_\textrm{\tiny{gr}}}$ and $T_{_\textrm{\tiny{gr}}}$ are given by (\ref{rhoVS}) and (\ref{Tgrav2}). In what follows we probe the fulfillment of (\ref{entprod}) in the models presented in Appendix~\ref{sec:nonconflat}. 

\subsection{Exponential functions}\label{expfuncts}

We consider only the solution (\ref{Uvars})-(\ref{NALB}) for $k_0=1$ and $f=\sin\chi$ which is the only physically viable case (its regularity is discussed in Appendix~\ref{sec:nonconflat}). The gravitational entropy density $16\pi \rho_{_\textrm{\tiny{gr}}}=|\Psi_2|$, the gravitational energy ${\cal E}_{_\textrm{\tiny{gr}}}= 16\pi \rho_{_\textrm{\tiny{gr}}}\,V$ and gravitational temperature defined in (\ref{rhoVS}) and (\ref{Tgrav2}) for this model follow readily from (\ref{NALB}), while the proper time rate of change $\dot {\cal E}_{_\textrm{\tiny{gr}}}$ and the gravitational temperature $T_{_\textrm{\tiny{gr}}}$ in (\ref{entprod}) are
\begin{eqnarray} \dot {\cal E}_{_\textrm{\tiny{gr}}}&=& \frac{L}{a\,N} \frac{\partial}{\partial \eta} (\rho_{_\textrm{\tiny{gr}}}\,V)=  \frac{\Delta^2\,f^4\,a_{,\eta}}{6\,a\,U^{\sqrt{2}}},\label{dotEgr1}\\
T_{_\textrm{\tiny{gr}}} &=& \left| \frac{(1+\sqrt{2})\Delta f^2}{\sqrt{2}\,U\,a}-\frac{a_{,\eta}}{U^{1+\sqrt{2}}\,a^2}\right],\label{Tgr1}\\
U &=& \exp\left(\frac{\Delta\,y(\chi)}{\sqrt{2}}\right),\quad y(\chi)=1-\cos\chi,\label{Udef}
\end{eqnarray}
Since the gravitational temperature $T_{_\textrm{\tiny{gr}}}$ is positive definite by definition and equation (\ref{dotEgr1}) shows that $ \dot {\cal E}_{_\textrm{\tiny{gr}}}$ is non-negative for an expanding model with $a_{,\eta}>0$, then the proper time rate of growth of the gravitational entropy $ \dot S_{_\textrm{\tiny{gr}}}$ in (\ref{entprod}) is also non-negative for an expanding model ($a_{,\eta}>0$):
\begin{equation}    \dot S_{_\textrm{\tiny{gr}}} = \frac{1}{T_{_\textrm{\tiny{gr}}}}\frac{L}{N}\left(\rho_{_\textrm{\tiny{gr}}}\,{\cal V}\right)_{,\eta}=\frac{\sqrt{2}(\sqrt{2}-1)\Delta^2\,f^4 U\,a\,a_{,\eta}}{6\left|\Delta f^2U^{\sqrt{2}}\,a-\sqrt{2}(\sqrt{2}-1)\,a_{,\eta}\right|}.\label{DotSthyp}\end{equation}
There is production of gravitational entropy as long as this scale factor is increasing with coordinate time, regardless of the physical assumptions used to determine $a(\eta)$. Notice that $f=0$ in the two symmetry centers and $ \dot {\cal E}_{_\textrm{\tiny{gr}}}= \dot S_{_\textrm{\tiny{gr}}}=0$, while $T_{_\textrm{\tiny{gr}}}$ becomes the FLRW expansion scalar $a_{,\eta}/a^2$.  

It is interesting to remark that (\ref{Tgr1}) and (\ref{DotSthyp}) also hold for the models with $k_0=0,-1$ despite the fact that the density in these cases becomes negative in a range comoving layers (see details in Appendix~\ref{sec:nonconflat}). 
 
\subsection{Sinusoidal functions}\label{sinusoidal}

The models described by the metric (\ref{metric3c})-(\ref{nu})  are ``closed'' (${\cal T}[t]$ diffeomorphic to $\mathbb{S}^3$). The gravitational entropy density $16\pi \rho_{_\textrm{\tiny{gr}}}=|\Psi_2|$, the gravitational energy ${\cal E}_{_\textrm{\tiny{gr}}}= 16\pi \rho_{_\textrm{\tiny{gr}}}\,V$ and gravitational temperature follow readily by computing (\ref{Egrav})-(\ref{Tgrav}) using \eqref{e0posvars1}-\eqref{e0posvars5}  
\begin{eqnarray} \dot {\cal E}_{_\textrm{\tiny{gr}}}&=& \frac{L}{a N} \frac{\partial}{\partial t} (\rho_{_\textrm{\tiny{gr}}}\,V)= \frac{\Delta^2\,\sin^4\chi\,[a_{,\eta}\,C_2+(a_{,\eta}\lambda -a\,\lambda_{,\eta})S_2]}{6\,a\,N\,L},\label{dotEgr2}\\
T_{_\textrm{\tiny{gr}}} &=&   \frac{\Delta\,\sin^2\chi\left |\left(LN_{,y}-\frac{1}{\sqrt{2}}L_{,y}N\right)-a_{,\eta}C_2-(\lambda a_{,\eta}-a\lambda_{,\eta})S_2\right|}{N}, \label{Tgr2}
\end{eqnarray}
where the functions $C_1,\,S_1,\,C_2,\,S_2,\,N,\,L$ are defined in (\ref{metric5b})  and non-negative factors were left out of the absolute value in (\ref{Tgr2}). 

Since $T_{_\textrm{\tiny{gr}}}$ is positive definite by construction, the condition for gravitational entropy production is already given by $\dot {\cal E}_{_\textrm{\tiny{gr}}}\geq 0$ in (\ref{dotEgr2}). However, we have now two time dependent functions $a$ and $\lambda$ ($\nu$ depends on $\lambda$, see (\ref{nu})), hence the conditions for gravitational entropy production $\dot S_{_\textrm{\tiny{gr}}}\geq 0$ are more complicated than those with exponential functions in (\ref{dotEgr1}) and (\ref{Tgr1}). We require now two conditions:
\begin{equation} a_{,\eta}\geq 0\quad \hbox{and}\quad \frac{a_{,\eta}}{a}-\frac{\lambda_{,\eta}}{\lambda}\geq 0, \label{cond_closed}\end{equation}
That is, for gravitational entropy production we require an expanding model with the FLRW-like scale factor $a$ growing at a faster rate than the free function $\lambda$. The expression for the proper time derivative $ \dot S_{_\textrm{\tiny{gr}}}$ follows from the quotient of (\ref{dotEgr2}) and (\ref{Tgr2})
\begin{equation}    \dot S_{_\textrm{\tiny{gr}}} = \frac{1}{T_{_\textrm{\tiny{gr}}}}\frac{L}{N}\left(\rho_{_\textrm{\tiny{gr}}}\,V\right)_{,\eta}=\frac{\Delta\,\sin^2\chi\,[a_{,\eta}\,C_2+(a_{,\eta}\lambda -a\,\lambda_{,\eta})S_2]}{6 a L\left |\left(LN_{,y}-\frac{1}{\sqrt{2}}L_{,y}N\right)-a_{,\eta}C_2-(\lambda a_{,\eta}-a\lambda_{,\eta})S_2\right|}.\label{DotStsin}\end{equation} 
Evidently, $\dot {\cal E}_{_\textrm{\tiny{gr}}}$ and $\dot S_{_\textrm{\tiny{gr}}}$ vanish and $T_{_\textrm{\tiny{gr}}}$ reduces to the FLRW expansion scalar at the symmetry centers $\chi=0,\pi$.  

To illustrate entropy production in the expanding ``closed'' models we examined in this section, we plot in Figure 1 the graphs of $\dot {\cal E}_{_\textrm{\tiny{gr}}},\,\,T_{_\textrm{\tiny{gr}}} $ and $\dot S_{_\textrm{\tiny{gr}}}$, as functions of $(a,\,\chi)$, obtained in (\ref{dotEgr2}), (\ref{Tgr2}) and (\ref{DotStsin}) and satisfying (\ref{cond_closed}), with the time evolution determined by (\ref{H0RW}) and (\ref{cuadratura1})-(\ref{cuadratura2}) and $\lambda = \lambda_0\,\hbox{sech}^2 a$ (a function rapidly decaying with increasing $a$). The choice of the free parameters is as follows $\Omega_M=0.3, \Omega_\Lambda = 0.7$ (see Section \ref{sec:timeevol}), $\Delta=0.2$. Notice the $\dot S_{_\textrm{\tiny{gr}}}$ function has the monotonically increasing expected behavior, as well as the energy and temperature functions being decreasing, where $T_{_\textrm{\tiny{gr}}}$ is monotonically decreasing, consistent with the end of structure formation and the dominance of the dark energy.  

\begin{figure}[H] 
\centering
\subfloat[\centering $\dot E_{_\textrm{\tiny{gr}}}$]{\includegraphics[width=0.48\textwidth]{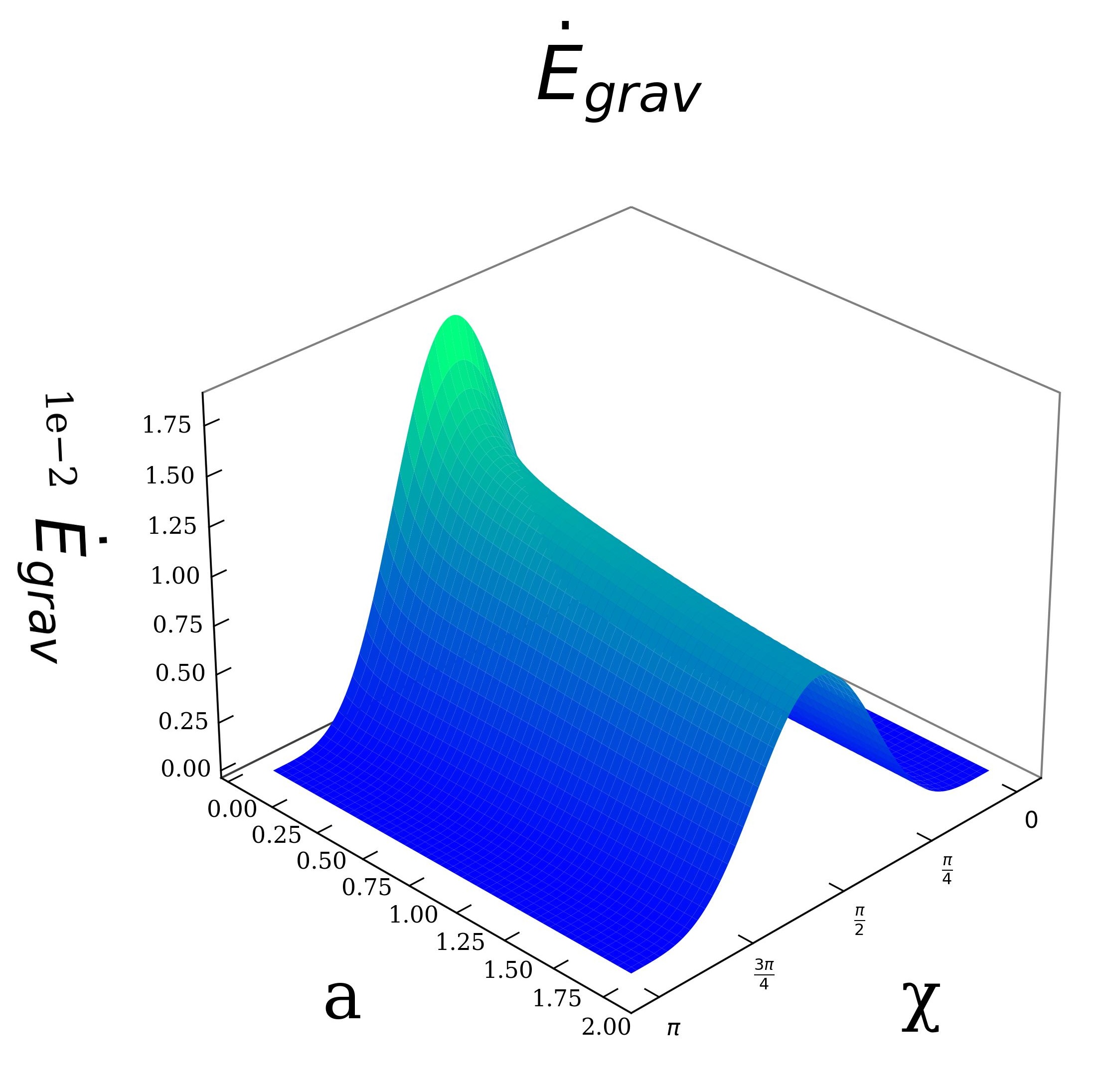}}
\hfill
\subfloat[\centering $T_{_\textrm{\tiny{gr}}}$]{\includegraphics[width=0.48\textwidth]{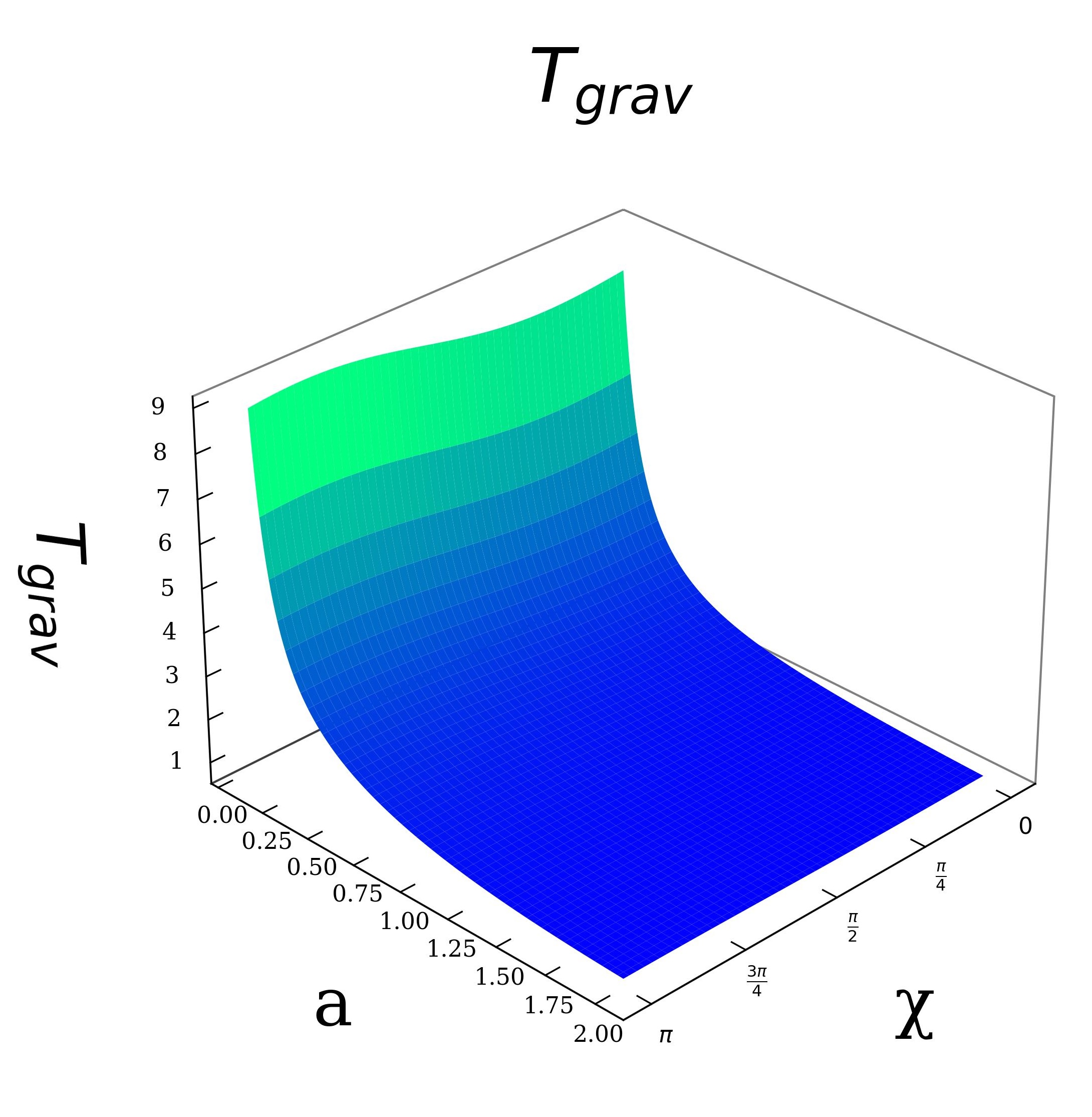}}\\
\subfloat[\centering $\dot S_{_\textrm{\tiny{gr}}}$]{\includegraphics[width=0.48\textwidth]{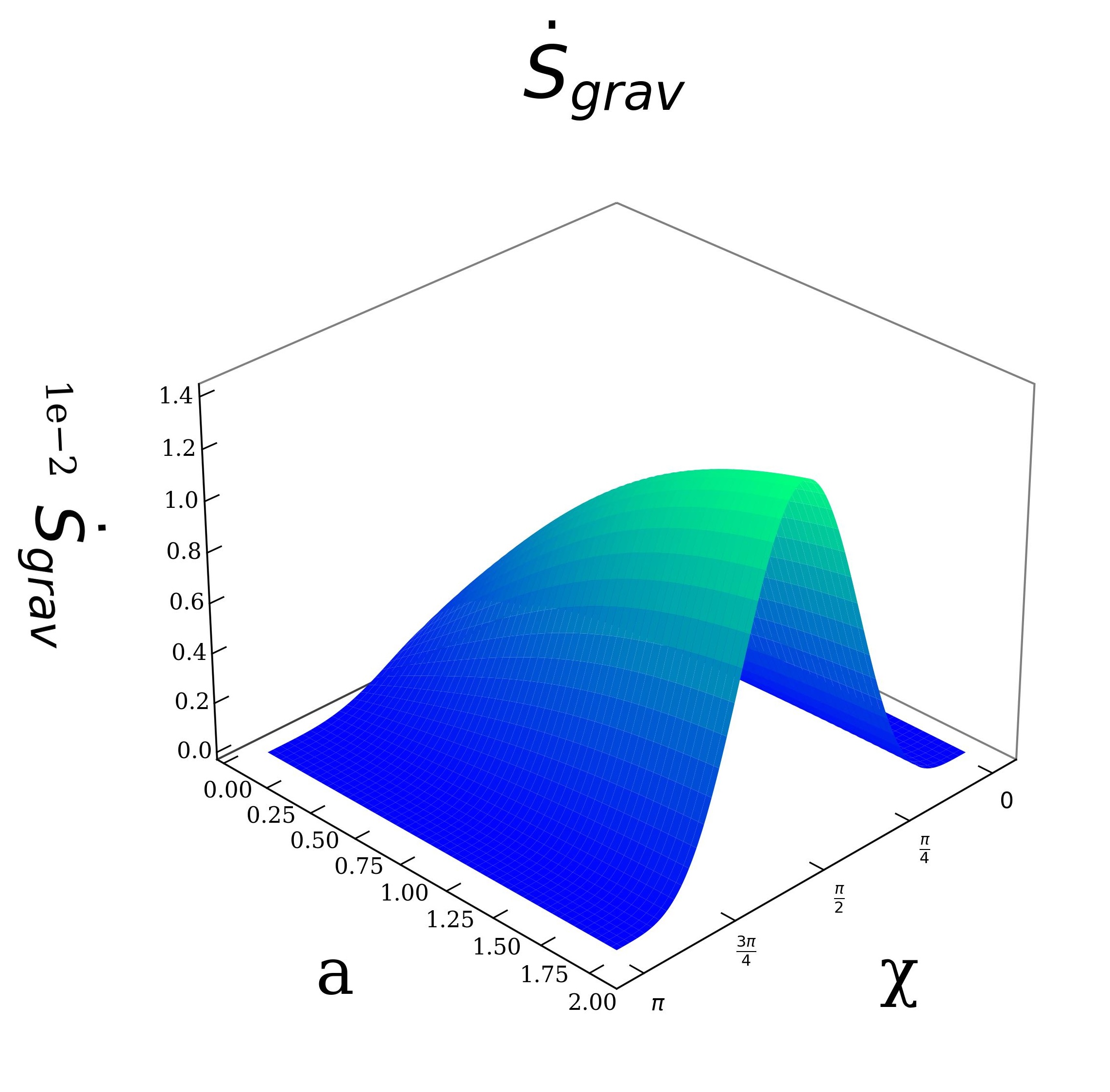}}
\caption{$\dot {\cal E}_{_\textrm{\tiny{gr}}},\,\,T_{_\textrm{\tiny{gr}}}$ and $\dot S_{_\textrm{\tiny{gr}}}$ as functions of $(a,\,\chi)$.}
\end{figure}
\section{Time evolution}\label{sec:timeevol}

While the spatial (radial) variation along spacelike slices ${\cal T}[t]$ is fully determined by the linear system (\ref{pdsNL}) (or its particular case (\ref{pdsNL2})), the time evolution is contained in 3 arbitrary functions $a,\,\lambda,\nu$ that emerge (once one function was used to redefine the time coordinate) as ``integration constants'' when integrating these linear systems.  

Ideally, we should determine these free functions by physical criteria (for example an equation of state), which would require separate dedicated articles. Since the function $a$ in (\ref{LN2}),(\ref{metric44}) and (\ref{metric3c}) is analogous to an FLRW scale factor (the FLRW limit is $N=\tilde L=1$), we propose to determine its time evolution by a Friedman equation of an FLRW model  with the same constant $k_0$, whose source is dust representing cold dark matter and a positive cosmological constant: 
 \begin{eqnarray} \left[\frac{\hat\Theta}{3 \hat H_0}\right]^2_{\textrm{RW}}&=&\frac{\hat H^2}{ \hat H_0^2}=\frac{a_{,\eta}^2}{a^4}= \frac{a_{,t}^2}{\hat H_0^2 a^2}=\frac{8\pi}{3 \hat H_0^2}\left(\frac{\hat \rho_0}{a^3}+\Lambda\right)-\frac{k_0}{\hat H_0^2 a^2}\nonumber\\&=&\frac{\hat\Omega_0^m-\hat\Omega_0^K\,a+\hat\Omega_0^\Lambda\,a^3}{a^3},\qquad  \hat\Omega_0^K=\frac{k_0}{\hat H_0^2}=\hat\Omega_0^m+\hat\Omega_0^\Lambda-1,\label{H0RW}
 \end{eqnarray} 
where $\hat\Theta$ is the FLRW expansion scalar and a hat $\hat{}$ will henceforth denote FLRW variables and parameters. Hence, $\hat H_0,\,\hat\Omega_0^m,\,\hat\Omega_0^K,\,\hat\Omega_0^\Lambda$ can be conceived as reference observational parameters of an analogous FLRW model. Equation (\ref{H0RW}) can also be used to eliminate the conformal time derivatives $a_{,\eta}$ and $a_{,\eta\eta}$ in terms of $a$ by
\begin{eqnarray}
 a_{,\eta} &=& \frac{a^2\hat H(a)}{\bar H_0}=\left[a(\hat \Omega_0^m-\hat\Omega_0^K a+\hat\Omega_0^\Lambda a^3)\right]^{1/2},\label{cuadratura1}\\ 
 a_{,\eta\eta}&=&a^2 \left[\left(\frac{\bar H}{\hat H_0}\right)_{,\eta}+2a\left(\frac{\hat H}{\hat H_0}\right)^2\right]=\frac12 \hat\Omega_0^m-\hat\Omega_0^K a+ 2\hat\Omega_0^\Lambda a^3, \label{cuadratura2}\end{eqnarray}
leading to the FLRW forms for density and pressure:
\begin{eqnarray}\frac{8\pi \hat \rho}{3\hat H_0^2}=\frac{1}{\hat H_0^2}\left[\frac{a_{,\eta}^2}{a^4} +\frac{k_0}{a^2}\right]=\frac{\hat\Omega_0^m}{a^3}+\hat\Omega_0^\Lambda,\qquad \frac{8\pi \hat p}{\hat H_0^2}=\frac{1}{\hat H_0^2}\left[\frac{a_{,\eta}^2}{a^4}-\frac{2 a_{,\eta\eta}}{a^3}-\frac{k_0}{a^2}\right]= -3\hat \Omega_0^\Lambda.\nonumber\\
\label{rhopFLRW}\end{eqnarray}
Equations (\ref{cuadratura1})-(\ref{cuadratura2}) and (\ref{rhopFLRW}) fully determine the time evolution of the variables in (\ref{Tht})-(\ref{eqp}) applied to the metric \eqref{metric44} of exponential solutions with separable metric coefficients that contains $a$ as the single time dependent free function. To fully determine the time evolution of the variables (\ref{Tht})-(\ref{eqp}) in the sinusoidal solution (for which $k_0=1$) described by the metric (\ref{metric3c})-(\ref{nu}),  we need to prescribe the other time dependent function as $\lambda=\lambda(a)$ (and $\nu$ through (\ref{nu})). We follow this process to plot the graphs of $\dot E_{_\textrm{\tiny{gr}}}, T_{_\textrm{\tiny{gr}}}$ and $\dot S_{_\textrm{\tiny{gr}}}$ in figure 1. 

\section{Compatibility with Penrose and Bonnor entropy criteria}\label{sec:arrow}

The notion of an ``arrow of time'' follows from Penrose's \cite{penrose1979singularities} qualitative proposal that the ratio of Weyl to Ricci curvature scalars is indicative of an evolution from an initial singularity towards later stages of structure formation along a suitable timelike direction. Assuming a suitable 4-velocity field parametrized by proper time, we have 
 \begin{equation} {\cal P} = \frac{C_{abcd}C^{abcd}}{R_{ab} R^{ab}} >1 \quad \hbox{as}\,\,\tau \,\,\hbox{increases with:}\quad \dot {\cal P}\geq 0,\label{ratioWR}\end{equation}
that should be a non-decreasing function. Based on this ratio Bonnor defined a gravitational entropy current as ${\cal P}^a = {\cal P}\,u^a$ and proved that condition (\ref{ratioWR}) holds for dust models (spherically symmetric LTB and Szekeres), but not for a shear-free heat conducting fluid such as the ones we have examined, but only when describing with them collapsing radiating spheres matched to a Vaidya exterior. 

Bonnor \cite{bonnor1985gravitational} and more recently Chkraborty et al  \cite{chakraborty2024arrow}  described the collapse of radiating  spheres by very simple particular cases of the models we have examined: the case $k_0=0,\,\,f=\chi$ of the separable metric (\ref{metric44}) in Sec.~\ref{separable}. These authors identified a ``time arrow'' as the future null direction along which the radiation in the Vaidya exterior increases, while the ``gravitational arrow'' was defined in terms of the evolution of (\ref{ratioWR}) as the sphere collapses. Bonnor found (confirmed by \cite{chakraborty2024arrow}) that the two ``arrows'' point in the opposite direction: as radiation increases, the ratio (\ref{ratioWR}) decreases with the collapse. 

However, Bonnor's result does not apply to the models we have examined. He identified the collapsing case as the only physically meaningful scenario for a heat conducting sphere matched to Vaidya, since an expanding sphere absorbing radiation from the Vaidya exterior would be unphysical.  Vaidya solution is characterized by the metric and coherent radiation energy momentum 
\begin{eqnarray}  ds^2&=& -\left(1-\frac{2m(u)}{r}\right)du^2 +2 du dr + r^2(d\theta^2+\sin^2\theta d\phi^2),\\
T^{ab} &=&\Phi(u)\, l^a l^b,\qquad \Phi= -\frac{1}{m}\frac{dm}{du}\end{eqnarray}
where $l^a$ is a null vector,  $m(u)$ and $\Phi(u)$ are the varying mass and radiating energy. Energy conditions require $\Phi\geq 0$, which implies a decaying mass $dm/du<0$ from the emission of radiation. The matching with Vaidya  places strong constraints on the CET entropy, since the conformal invariant $\Psi_2$  for the Vaidya solution is   
\begin{eqnarray}   \Psi_2 = \frac{m(u)}{r^3}\quad\Rightarrow\quad  16 \pi \rho_{_\textrm{\tiny{gr}}}= \frac{16\pi |m(u)|}{r^3},\quad 
 \Rightarrow\quad {\cal E}_{_\textrm{\tiny{gr}}}=16\pi\rho_{_\textrm{\tiny{gr}}}\,r^3=16\pi |m(u)|,\end{eqnarray}
hence, since energy conditions require  a decreasing mass $dm(u)/du<0$, then the gravitational energy ${\cal E} _{_\textrm{\tiny{gr}}}$ is positive, but must also decrease, which necessarily leads also to a decreasing gravitational entropy $\dot S_{_\textrm{\tiny{gr}}}=
\dot {\cal E}_{_\textrm{\tiny{gr}}}/T_{_\textrm{\tiny{gr}}}<0$. 

As a contrast with collapsing spheres in a Vaidya exterior, we have examined more general cases of the same class of shear-free heat conducting solutions, but as expanding cosmological models not restricted by any matching. In Bonnor's example, the only time dependent free function is completely determined by the matching with the Vaidya exterior, while as we showed in Section~\ref{sec:timeevol} the equivalent time dependent free function can be arbitrarily defined, but (as we argued) the best choice in a cosmological context is to define it as a FLRW scale factor, with the resulting models approximating FLRW models without the need to perform a smooth matching with them. In what follows we show that the expanding models we have examined comply with (\ref{ratioWR}). 

For the models under consideration, we have $C_{abcd}C^{abcd} = 8E_{ab}E^{ab}=8\Psi_2^2$, therefore, we can write (\ref{ratioWR}) as
\begin{equation}\frac{8E_{ab}E^{ab}}{R_{ab}R^{ab}}= \frac{\frac43 \Psi_2^2}{8\pi (\rho^2+3p^2-2q_aq^a) },\qquad \Psi_2 =\frac{\Delta^2\, f^2\,\tilde L^2}{6\,a^2}.\label{ratioWR2}\end{equation}
To evaluate this quotient we consider only the leading terms of series expansions of $\rho,\,p,\,q_a$ on the free constant parameter $\Delta$ that appears in the functional forms of these quantities for the metric \eqref{metric} in \eqref{eqQapp}-\eqref{eqp}. Since $\Delta$ always appears as the product $\Delta\,y$, the proposed  expansion is only appropriate for the case $k_0$ for which $y$ is bounded ($0\leq y\leq 2$). The expansions are easily computed by applying \eqref{eqQapp}-\eqref{eqp} to $L$ and $N$ given by the metrics \eqref{metric44} and \eqref{metric3c}.  The leading terms of the expansion of (\ref{ratioWR2}) provide a comparative  context with cosmological perturbatios of FLRW spacetimes in a comoving gauge. 

\begin{figure}[H]
\centering
\subfloat[\centering]{\includegraphics[width=0.485\textwidth]{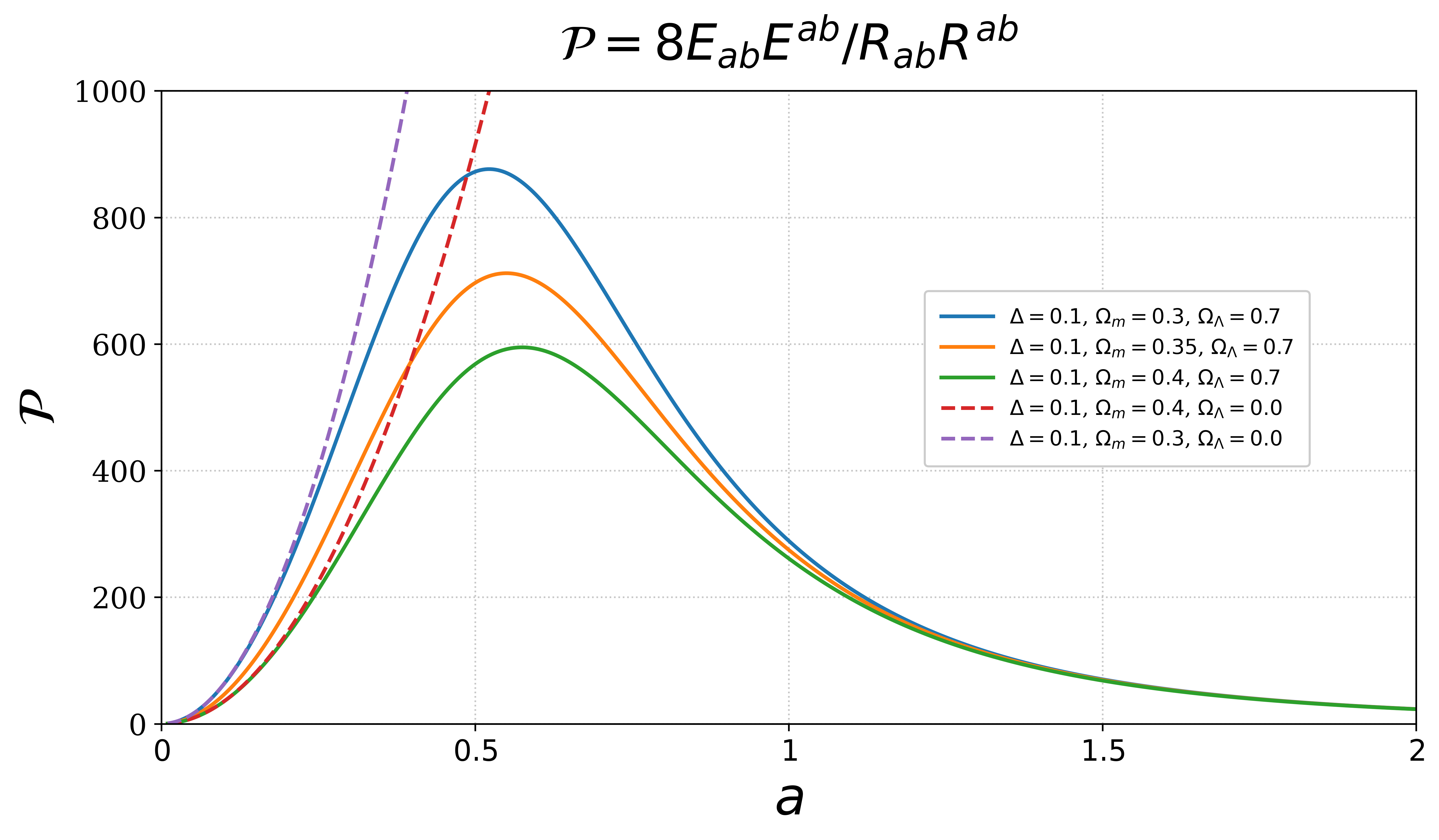}\label{fig:Omegaplot}}
\hfill
\subfloat[\centering]{\includegraphics[width=0.485\textwidth]{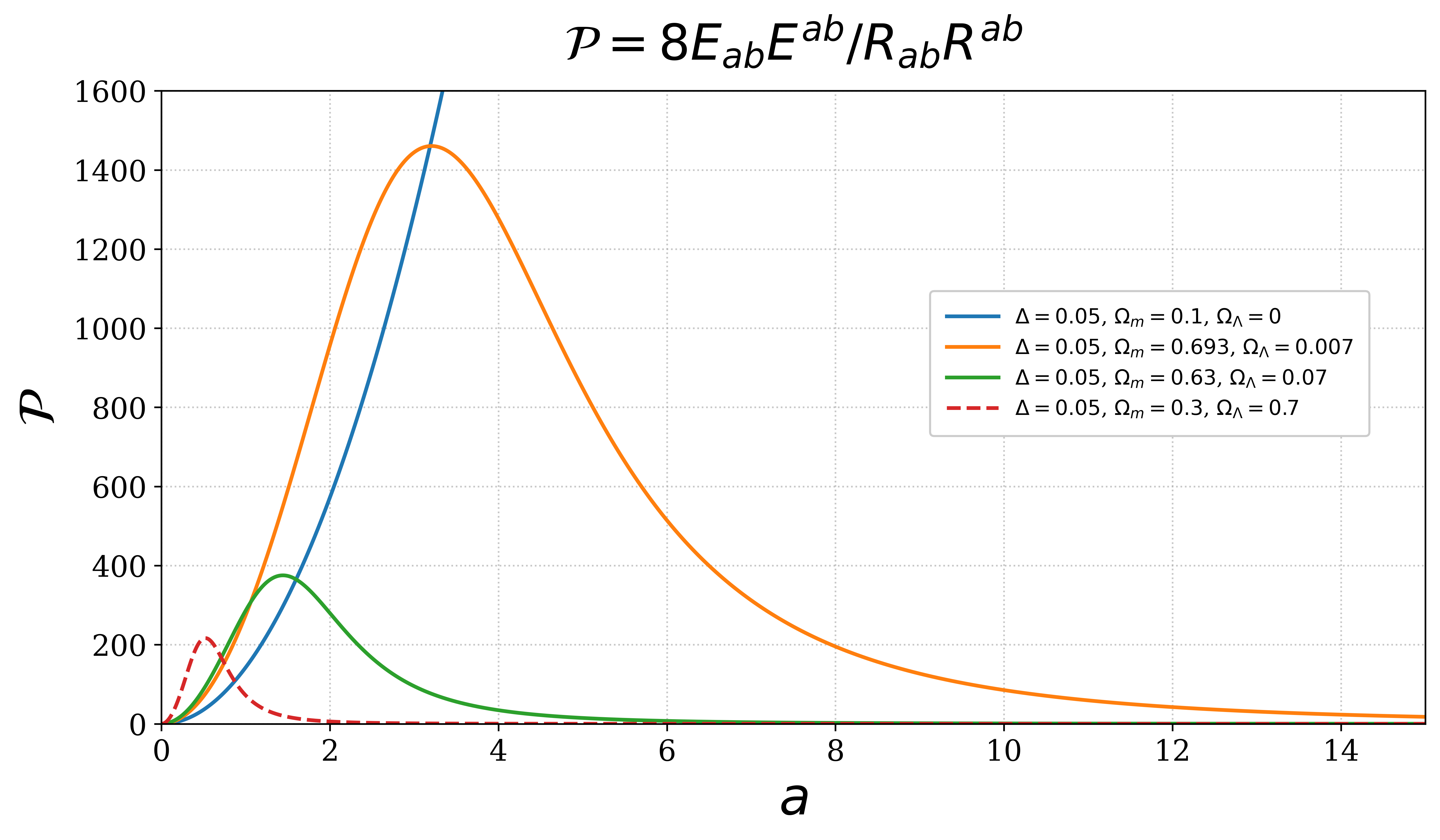}\label{fig:Lambdaplot}}
\caption{Figure (a) shows radial profiles for the $\mathcal{P}$ entropy current function with different values of $\Omega_M$. Figure (b) shows radial profiles for the $\mathcal{P}$ entropy current function with different values of $\Omega_\Lambda$. Refer to the main text.}
\end{figure}

Considering (\ref{H0RW}) and (\ref{rhopFLRW}), the leading terms of the power expansions of \ref{ratioWR2} on $\Delta$ (which are used for the plots shown in Fig.~2. are)
 \begin{eqnarray} \frac{8\pi\hat \rho}{3 H_0^2} &=&   \frac{\hat\Omega_0^m}{a^3}+\Omega_0^\Lambda,\qquad \frac{8\pi\hat p}{H_0^2} = -3\Omega_0^\Lambda,\\
 \Psi_2 &=& \frac{\sin\chi}{6 a^2}\Delta^2,\qquad \frac{64\pi^2 q_aq^a}{\hat H_0^4} = \frac{2(1+\sqrt{2})^2}{\hat H_0^2}\frac{\Omega_0^m-\Omega_0^K+\Omega_0^\Lambda}{a^5}\Delta^2,\end{eqnarray}
Fig.~\ref{fig:Omegaplot} shows the radial profiles (evaluated at $\chi=\pi/2)$ of Bonnor's gravitational entropy current function, $\mathcal{P}$. Notice that when considering dark energy modeled by a cosmological constant $\Lambda$ there is a decay of $\mathcal{P}$ instead of the monotonous growth when there is no cosmological constant, see Fig.~\ref{fig:Lambdaplot}. Similar results have been reported in the literature \cite{gron2002weyl, gron2012entropy, marozzi2015cosmological, sussman2014gravitational} this effect could be attributed to the accelerated expansion from $\Lambda$ which affects structure formation.

\section{Conclusion and further work}\label{sec:conclusions}

We have examined the growth of gravitational entropy in the framework of the gravitational entropy proposal of Clifton, Ellis and Tavakol (CET), applied to  the spherically symmetric case of exact solutions of Einstein's equations characterized by an irrotational  shear-free fluid admitting energy flux in a comoving frame, which in general no not admit isometries. We consider these solutions within a cosmological context, as opposed to previous literature using them as  toy models of radiating spheres collapsing in a Vaidya background. 

Besides probing the CET proposal, we have also examined the connection between the dynamical equations of this gravitational entropy proposal and Einstein's equations in the 1+3  fluid flow formulation. As a result, we are able to relate the variables of the gravitational entropy with the covariant objects that generate the inhomogeneity of the solutions (Weyl tensor and 4-acceleration and heat flux vectors).  We found for  models with both open and closed hypersurfaces orthogonal to the 4-velocity specific examples fulfilling energy and regularity conditions, with the potential  to describe tractable inhomogeneous and anisotropic generalizations of FLRW models that can approximate them in a smooth and controllable form. 

We also verified that when models that comply with regularity conditions undergo cosmic expansion they comply in general with gravitational entropy growth in the framework of the CET proposal. They also comply with the notion of a gravitational ``arrow of time'' proposed by Penrose that relates structure formation to a non-decreasing ratio of Weyl to Ricci scalar curvature, though our results show a rapid decay of this ratio associated with the cosmological constant $\Lambda$, a finding that is consistent with the slowing of the growth structure formation in a dark energy scenario. These results stand in contrast with those of previous work showing a decrease of the ratio of Weyl to Ricci curvature with  the same class of solutions, but regarded as toy models of collapsing spheres in a Vaidya exterior.

The results of the present article show that the models we have examined have a good potential to be useful in cosmological applications. To develop this potential for these models, we need to examine the compatibility of their source (a fluid with energy flux) with physical assumptions that can be useful in cosmology. In particular, for considering cosmological applications we need to drop the interpretation of their energy flux as heat conduction (as in their usage as radiating spheres), as this physical model is only appropriate for a hydrodynamical thermal regime based on very short range interactions. A much more appealing physical interpretation of the energy flux is in terms of non-comoving peculiar velocities in a non-relativistic limit. In previous work we have tested this interpretation on Szekeres models of class II, which also admit a non-zero energy flux \cite{najera2020pancakes, najera2021non}. It is our purpose for future work to follow similar lines with the present models, which have more degrees of freedom and are technically more challenging, all of which have the potential to describe more realistic peculiar velocity fields.

\bmhead{Acknowledgements}
SN acknowledges support form Postdoctoral grant from Secretar\'ia de Ciencia, Humanidades, Tecnolog\'ia e Innovaci\'on SECIHTI. JCH acknowledges support from the UNAM-PAPIIT grant IG102123 "Laboratorio de Modelos y Datos (LAMOD) para proyectos de Investigaci\'on Cient\'ifica: Censos Astrof\'isicos", as well as from SECIHTI (formerly CONAHCYT) grant CBF2023-2024-162 and DGAPA-PAPIIT-UNAM grant IN110325 "Estudios en cosmolog\'ia inflacionaria, agujeros negros primordiales y energ\'ia oscura."

\begin{appendices}

\section{Derivation of shear-free solutions with energy flux}\label{sec:shearfreeapp}

Practically all previous publications on spherically symmetric shear-free solutions admitting a nonzero energy flux have used  the case $k_0=0$ in the following metric in a comoving frame
\begin{eqnarray}
ds^2 &=& -A^2 dt^2+B^2\left[\frac{dr^2+ r^2(d\theta^2+\sin^2\theta d\phi^2)}{\left(1+\frac14k_0r^2\right)^2}\right],\label{metricA1}
\end{eqnarray}
 This metric becomes metric \eqref{metric}  (which we repeat below)
 \begin{eqnarray}
ds^2 &=& \frac{-N^2 dt^2+d\chi^2+ f^2(\chi)(d\theta^2+\sin^2\theta d\phi^2)}{L^2},\label{metricApp}\\
f(\chi)&=&\chi,\,\sin\chi,\,\sinh\chi\quad \hbox{for}\quad k_0=0,1,-1,\nonumber
\end{eqnarray} 
by redefining the metric functions as $A=N/L,\,\,B=1/L$ and performing the following coordinate transformation   
\begin{eqnarray}\chi = \int_0^r{\frac{dr}{1+\frac14k_0r^2}} = \left\{  \begin{array}{ll} 
          r & k_0=0 \\
          2\arctan (r/2)& k_0=1 \\
          2\hbox{arctanh}(r/2) & k_0=-1 
    \end{array}\right.,\label{chi}\end{eqnarray}
 %
%
%
 %
 %
To obtain exact solutions we use the following constraint that folows from the condition for pressure isotropy: $G^\chi_\chi-G^\theta_\theta=0$ computed with \eqref{metricApp} 
\begin{equation} \left(\frac{2L''}{L}-\frac{N''}{N}\right)\,f+\left(\frac{N'}{N}+\frac{L'}{L}\right)\,f' =0,\label{CPI1}\end{equation}
where the primes denote  $\partial/\partial \chi$. The linear system \eqref{pdsNL} follows by introducing the variable $y$ 
%
%
%
\begin{eqnarray}  f^2(\chi) &=& y(2-k_0 y)\quad\Rightarrow\quad 
      y(\chi)= \left\{  \begin{array}{ll} 
         \chi^2/2,& k_0=0,  \\
         1-\cos\,\chi,& k_0=1, \\
         \cosh\,\chi-1, & k_0=-1.  
       \end{array}\right.\label{ychiapp}
\end{eqnarray}       
The following equations are useful to relate $\chi$ and $y$
\begin{eqnarray}
\frac{\partial}{\partial \chi}&=&y'\frac{\partial}{\partial y}=f\frac{\partial}{\partial y}=\sqrt{y(2-k_0y)}\frac{\partial}{\partial y},\label{transf2}\\
y'&=&\frac{dy}{d\chi}=f(\chi),\qquad N^\prime=\frac{ \partial N}{\partial \chi} =  f(\chi)\frac{ \partial N}{\partial y},\qquad L^\prime=\frac{ \partial L}{\partial \chi} =  f(\chi)\frac{ \partial L}{\partial y},\label{gradsNL}
\end{eqnarray}
Since the system \eqref{pdsNL} allows for the elimination of second order radial derivatives in terms of $N$ and $L$, we obtain for \eqref{metricApp} and energy momentum tensor \eqref{Tab}  the following forms for the kinematic parameters: expansion $\Theta=u^a\,_{;a}$, 4-acceleration $\dot u_a =u_{a;b}u^b$, the electric Weyl tensor $E^a_b$ (the magnetic Weyl tensor vanishes), energy flux $q_a$, density $\rho$ and pressure
\begin{eqnarray}
\Theta &=& -\frac{3\dot L}{L}=-\frac{3L_{,t}}{N}\, \label{Tht}\\
 \dot u_a &=& {\cal A}\delta_a^\chi,\quad {\cal A} = \frac{N'}{N}-\frac{L'}{L},\label{Accapp}\\
E^a_b&=&\hbox{diag}[0,-2\Psi_2,\Psi_2,\Psi_2],\quad \Psi_2 = -\frac{L^2 f^2\,J}{6}, \label{EWeylapp}\\
8\pi q_a &=& 8\pi Q\delta_a^\chi,\quad Q= -\frac23\Theta' =-\frac{2L'_{,t}}{N}+\frac{2N' L_{,t}}{N^2}\label{eqQapp}\\
\frac83\pi\rho &=& \left(\frac{\Theta}{3}\right)^2+k_0 L^2+2LL'\,ff'+\left(L'^2-\frac23 JL^2\right)\,f^2\label{eqrho}\\
8\pi p &=&  -\frac{\Theta^2}{3}-\frac23\dot \Theta-k_0 L^2+2\left(\frac{N'}{N}-\frac{2L'}{L}\right)L^2\,ff'
+ \left(\frac{3L'^2}{L^2}-\frac{2L'\,N'}{L\,N}\right) L^2\,f^2,\nonumber\\
\label{eqp}
\end{eqnarray}
while energy-momentum balance $T^{ab}\,_{;b}=0$ leads to
\begin{eqnarray}
&{}&\dot\rho-3(\rho+p)\frac{\dot L}{L}+\left[\left(Q'+\left(\frac{2N'}{N}+\frac{3L'}{L}\right)Q\right)f^2+3Qf'\right]\,L=0,\label{EMC1}\\
&{}&\left(\dot Q-\frac{3\dot L}{L}Q\right)L+p'+\left(\frac{N'}{N}-\frac{L'}{L}\right)(\rho+p)=0,\label{EMC2}
\end{eqnarray}
where the ``dot'' denotes proper time derivative for the 4-velocity $u^a=\sqrt{-g^{tt}}\delta^a_t=(L/N)\delta^a$, while $\Psi_2$ is the only nonzero conformal invariant (the magnetic Weyl tensor is zero, hence solutions are Petrov type D or O if $J=0\,\,\Rightarrow\,\,\Psi_2=0$).

\section{Non-conformally flat solutions}\label{sec:nonconflat}

\subsection{Exponential solutions}\label{exponential}

Assuming that $\lambda,\,\nu$ are both nonzero in the metric \eqref{LN2}, it is convenient to rewrite the solutions (\ref{epos}) as in (\ref{LN}) but in terms of a single exponential function, leading to the following metric functions 
\begin{eqnarray} N= \frac{\beta +U^{2\sqrt{2}}}{U^{\sqrt{2}}},\qquad L=\frac{\beta+ U^2}{U},\qquad U = \exp\left(\frac{\Delta\,y(\chi)}{\sqrt{2}}\right),\label{Uvars}\end{eqnarray} 
where $\beta=\lambda=\nu$ in \eqref{LN2} follows from the regularity condition \eqref{reg1} (see Appendix \ref{regularity}) at the symmetry center as $\chi=0$ (since $U(0)=1$). In what follows we examine the different parameter cases associated with (\ref{Uvars}).
\subsubsection{The general case: two time dependent free functions}\label{open}

In the case with $\beta\ne 0$ the coordinate $\chi$ (and thus $y(\chi)$) can reach infinite values. However, we will not consider these cases, since as proved in Appendix \ref{regularity} they exhibit a singular second symmetry center as $\chi\to\infty$.  The case $k_0=1$  avoids the problematic limit $\chi\to\infty$, but the regularity condition (\ref{reg1}) at the second symmetry center at $\chi=\pi$ implies $\beta$ constant. However, numerical trials with  (\ref{Tht})-(\ref{eqp}) show that pressure takes too large negative values in the allowed coordinate range because of terms proportional to $f$ and $f'$ in (\ref{eqp}). 

\subsubsection{One free function: separable metric functions}\label{separable}

Particular solutions with $\beta=0$ in \eqref{Uvars} leads to the metric \eqref{LN2} with separable metric functions, leading to the particular case of  
 \begin{eqnarray} ds^2=\frac{a^2(\eta)}{L^2_{(A)}(\chi)}\left[-N^2_{(B)}(\chi) d\eta^2 + d\chi^2+f^2(\chi)(d\theta^2+\sin^2\theta d\varphi^2\right],\label{metric44}\end{eqnarray}
 where $(A),(B)=1,2$ denoting any one of the 4 combinations that follow from the functions
\begin{eqnarray} N_1 =U^{\sqrt{2}}\quad N_2 = U^{-\sqrt{2}},\qquad L_1 = U,\quad L_2 = U^{-1},\label{NALB}\end{eqnarray}
where $U$ is defined in (\ref{Uvars}) for the three cases of $y(\chi)$ in (\ref{ychi}). 

In the cases $k_0=0,-1$ we found that $N_1$ and $L_2$ is the only combination in (\ref{NALB}) that is compatible with a regular radial asymptotic regime. This case was examined in Section \ref{expfuncts}, it complies with gravitational entropy growth. However,  it is physically viable only for values of $\chi$ close to the symmetry center, since $\rho$ necessarily becomes negative for large values of $\chi$. This follows from substituting  $N=N_1$ and $L=L_2/a$ (with $L_,t =a\,L_{,\eta}$)  in \eqref{eqrho}  
\begin{equation}\frac{8\pi}{3} \rho = \frac{a_{,\eta}^2}{a^4}\,\frac{1}{U^{2(1+\sqrt{2})}}
 +\frac{k_0-\sqrt{2}\Delta f'-\frac16 \Delta^2 f^2}{ a^2\,U^2}\end{equation}
where we notice that the second term in the right hand side is negative and dominates the first term for large values of $\chi$. In the case $k_0=1$ we have $0\leq \chi\leq \pi$ and so the exponential functions in (\ref{NALB}) are positive. However,  the regularity condition \eqref{reg1} does not hold in the second symmetry center at $\chi=\pi$ (since $U(\pi)=\exp(\sqrt{2}\Delta)>1$).

\subsection{Sinusoidal solutions}\label{sinusoidal} 

The exact sinusoidal solutions of (\ref{pdsNL2}) with $e_0=-1$ in (\ref{eneg}) admit a function $y(\chi)$ given by the three options of (\ref{ychi}). However, we keep only the option $y(\chi)=1-\cos\chi$ for the case $k_0=1$, since in the cases $k_0=0,-1$ a periodic sinusoidal function of the form $\sin(\chi^2/2)$ or $\sin(\cosh \chi-1)$ will necessarily exhibit increasingly rapid oscillations as $\chi$ increases. 

The metric \eqref{metricApp} takes the form 
 \begin{eqnarray} ds^2 = a^2(\eta) \left[\frac{-\left[C_1+\nu\,S_1\right]^2 d\eta^2+d\chi^2+ \sin^2(\chi)(d\theta^2+\sin^2\theta d\phi^2)}{\left[C_2+\lambda\,S_2\right]^2}\right].\label{metric3c}\end{eqnarray}
 \begin{eqnarray}
 C_1&=&\cos \Delta y(\chi),\; S_1 = \sin \Delta y(\chi),\; C_2=\cos \frac{\Delta y(\chi)}{\sqrt{2}},\; S_2=\sin \frac{\Delta y(\chi)}{\sqrt{2}},\label{metric5b}
 \end{eqnarray}
 where $y(\chi)=1-\cos\chi$. The  solutions admit two symmetry centers marked by $\chi = 0,\pi$ (or $y=0,2$). The metric (\ref{metric3c}) fulfills regularity condition (\ref{reg1}) at $\chi=0$ (since $C_1(0)=C_2(0)=1$ and $S_1(0)=S_2(0)=0$), but imposing this condition on the second symmetry center at $\chi=\pi$ implies the following constraint between $\lambda$ and $\nu$:
\begin{equation}\nu = n_1\,\lambda +n_2,\qquad n_1 = \frac{\sin \sqrt{2}\Delta}{\sin 2\Delta},\quad n_2 = \frac{\cos \sqrt{2}\Delta -\cos 2 \Delta}{\sin 2\Delta},\label{nu}\end{equation}
which leaves the most general case with two time dependent free functions $a(\eta)$ and $\lambda(\eta)$. 
Geometric and physical variables follow from \eqref{Tht}-\eqref{eqp} specialized to the metric (\ref{metric3c}
 \begin{eqnarray}
 \frac{\Theta}{3} &=& \frac{a_{,\eta}\, L-\lambda_{,\eta}\,a\,S_2}{a^2\,N},\qquad  \Psi_2 = \frac{\Delta^2 \sin^2\chi\, L^2}{6a^2}, \label{e0posvars1}\\
  {\cal A} &=& \left(-\frac{L_{,\chi}}{L}+\frac{N_{,\chi}}{N}\right)=f\,\Delta \left(\frac{\nu C_1-S_1}{N}-\frac{\lambda\, C_2 -S_2}{\sqrt{2}L}\right)\label{e0posvars2}\\
 4\pi Q &=& -\frac{1}{3}\mathcal{A}\Theta -\frac{1}{\sqrt{2}aLN}f\,\Delta \, \lambda_{,\eta}, \label{e0posvars3}\\
  \frac{8\pi}{3} \rho &=&  \left(\frac{\Theta}{3}\right)^2+\left(1-\frac{\Delta^2\sin^2\chi}{3} \right)\,\frac{L^2}{a^2} -\frac{L_{,\chi}}{a^2\sin\chi }\,\left[L_{,\chi}\sin\chi-2L\cos\chi \right],\label{e0posvars4}\\
 8\pi p&=& -\frac{\Theta^2}{3}-\frac23 \dot \Theta +\left[-1+\left(\frac{3L_{,\chi}}{L}-\frac{2N_{,\chi}}{N}\right)\frac{ L_{,\chi}}{L}+2\left(\frac{N_{,\chi}}{N}-\frac{2L_{,\chi}}{L}\right)\frac{\cos\chi}{\sin\chi}\right]\,\frac{L^2}{a^2},\nonumber\\ 
\label{e0posvars5}
 \end{eqnarray}
where $\lambda$ and -\eqref{nu} are related by \eqref{nu}.
 
 \section{Regularity conditions}\label{regularity}

The metric functions in \eqref{metricApp} must be compatible with a regular symmetry center at $\chi=0$ (worldline that marks fixed points of the SO(3) group), so that the metric functions $N,\,L$ and any scalar function $\Phi$ must comply with
\begin{eqnarray}
\lim_{\chi\to 0}\frac{N(t,\chi)}{L(t,\chi)}=1,\qquad \lim_{\chi\to 0} \frac{\partial \Phi(t,\chi)}{\partial\chi}=0,\label{reg1}
\end{eqnarray}
with the same limits holding also at $\chi=\pi$ for the case $k_0=1$. Also, the metric functions $N,\,L$ must be regular throughout the range of $\chi$, in particular in the cases $k_0=0,-1$ whose coordinate range can reach $\chi\to \infty$.

The resemblance of the metric \eqref{metricApp} to an FLRW metric might be a coordinate effect that does not (necessarily) imply shared coordinate free geometric properties.  The difference between models with metric \eqref{metricApp} and FLRW geometry can be appreciated through the proper length $\ell=\int{\sqrt{g_{\chi\chi}}d\chi}$ of radial rays $[t_0,\chi,\theta_0,\varphi_0]$ that are geodesics ($ds^2=d\ell^2=g_{\chi\chi}d\chi^2$) of the 3-dimensional hypersurfaces  ${\cal T}[t]$ orthogonal to the 4-velocity.  

In FLRW metrics ($N=L=1$) the radial coordinate determines also the proper length $\ell=\int{\sqrt{g_{\chi\chi}}d\chi}=\chi$ along radial rays. However, in the inhomogeneous models under consideration the metric coefficient $g_{\chi\chi}$ in \eqref{metricApp} explicitly depends on $\chi$, the range of the coordinate $\chi$ in (\ref{ychiapp}) might significantly differ from the range of the proper length $\ell$: 
\begin{eqnarray}k_0&=& 0,-1,\qquad \lim_{x\to\infty}\ell(\chi)= \lim_{\chi\to\infty}\int_0^\chi{\sqrt{g_{\chi\chi}}d\chi}=\lim_{\chi\to\infty}\int_0^\chi{\frac{d\chi}{L(\eta,\chi)}},\label{ell1}\\
k_0&=& 1,\; 0\leq \chi\leq \pi:\quad \lim_{\chi\to\pi}\ell= \lim_{x\to\pi}\int_0^\pi{\sqrt{g_{\chi\chi}}d\chi}=\lim_{\chi\to\pi}\int_0^\chi{\frac{d\chi}{L(\eta,\chi)}},\label{ell2}
\end{eqnarray}
The limit \eqref{ell1} is always infinity in FLRW metrics, but might be finite in \eqref{metricApp} if $\lim_{\chi\to\infty} L=0$, which might happen in solutions of the linear system \eqref{pdsNL}. Likewise, while the limit (\ref{ell2}) is always finite ($\ell=\pi$) in FLRW metrics, it might be infinity in \eqref{metricApp} if $L$ has a zero for a finite $0<\chi^*<\pi$. In these cases the time slices are different from those of an FLRW model with same $k_0$. However, the time slices are qualitatively analogous to those of FLRW geometry if $L>0$ and is bounded in the full coordinate range of $\chi$. 

It is easy to illustrate how different from an ``open'' FLRW model (with $k_0=0,-1,\,f=\chi,\,\sinh\chi$)  can be an inhomogeneous model with metric \eqref{metricApp} and same $k_0,\,f(\chi)$. As an example,  the solutions \eqref{Uvars}  with an infinite  range of the coordinate $\chi$ ($k_0=0,-1$) have a finite proper radial length $\ell$ along radial rays as $\chi\to\infty$, so it is ``closed'' model that exhibits an unsuspected singularity. The proper asymptotic behavior of metric coefficients and radial proper length are
\begin{eqnarray}
\lim_{\chi\to\infty} \sqrt{-g_{\eta\eta}}&=&a\, \lim_{\chi\to\infty}  \frac{\beta +U^{2\sqrt{2}}}{(\beta+U^2)U^{\sqrt{2}-1}}=\infty,\qquad \lim_{\chi\to\infty} \sqrt{g_{\chi\chi}}=a\,\lim_{\chi\to\infty} \frac{U}{\beta+U^2}=0,\nonumber\\
\lim_{\chi\to\infty}\ell(\chi) &=& a \lim_{\chi\to\infty}\int_0^\chi{\frac{U\,d\chi}{\beta+U^2}}=\hbox{finite},\qquad  \lim_{\chi\to\infty}\sqrt{g_{\theta\theta}} =  a\lim_{\chi\to\infty}\frac{U\,f}{[\beta+U^2]}=0,\nonumber\\\label{limits}\end{eqnarray}
These limits show that the hypersurfaces ${\cal T}[t]$ admit a second symmetry center at a finite proper length along the ${\cal T}[t]$, even though they are marked by infinite values of the radial coordinate $\chi$.  It is straightforward to show that this second symmetry center marks a curvature singularity, as simple inspection of (\ref{EWeylapp})-(\ref{eqp}) shows that the metric functions and their radial gradients that determine Weyl and Ricci curvature ($\Psi_2,\,\rho$ and $p$) diverge in the limit $\chi\to\infty$ as $\sim L^2\sim U^2$, even if quotients like $L'/L,\,N'/N,\, L_{,t}/N,\,L/N$ converge. 

\section{1+3 Equations}\label{sec:1p3eq}
Einstein's equations are equivalent to the first order 1+3 fluid flow equations. For the models under consideration these equations are restricted by $\sigma_{ab}=\pi_{ab}=H_{ab}=\omega_{ab}=0$ 
\begin{itemize}
\item\textbf{Evolution equations}
\begin{eqnarray}
\dot{\rho} &=& - (\rho + p)\Theta - \nabla_a q^a -2 \dot{u}_a q^a, \\
\dot{\Theta} &=&- \frac{1}{3}\Theta^2 - 4\pi(\rho+3p+2\Lambda) + \nabla_a \dot{u}^a 
 +\dot{u}_a \dot{u}^a, \\
\dot{q}_{\langle a\rangle} &=& - \frac{4\pi}{3}\Theta q_a - (\rho+p)\dot{u}_a - \nabla_a p, \\
\dot{E}_{\langle ab\rangle} &=&- \Theta E_{ab} -4\pi ( \nabla_{\langle a}q_{b\rangle} 
   +2 \dot{u}_{\langle a}q_{b\rangle}), .
\end{eqnarray}

\item\textbf{Constraints:}
\begin{eqnarray}
E_{ab} &=& \nabla_{\langle a}\dot{u}_{b\rangle} 
  \dot{u}_{\langle a}\dot{u}_{b\rangle}, \\
 \frac{2}{3}\nabla_a \Theta &=& 8\pi q_a,\\
\nabla^b E_{ab} &=&- \frac{4\pi}{3}(2\Theta q_a+ 2\nabla_a \rho )\\
{}^3R &=& 16\pi \rho - \frac{2}{3}\Theta^2.
\end{eqnarray}
\end{itemize}

The specialization of these equations to the solutions under consideration leads to
\begin{eqnarray}
\dot\Theta &=& - \frac{\Theta^2}{3} 
- 4\pi (\rho + 3p-2\Lambda) -\left(-\mathcal{A}_{,\chi} + \frac{2 \mathcal{A} f_{,\chi}}{f} + \mathcal{A}^2 \right) L^2, \\
\dot\rho &=& - \left(Q_{,\chi} - \frac{2 Q f_{,\chi}}{f} - 2 Q \mathcal{A}\right) L^2 \, - (\rho + p)\,\Theta, \\
\dot Q &=& \frac{4}{3}\Theta Q - \mathcal{A}(\rho+p) - p_{,\chi}, \\
\dot \Psi_2 &=& - \Theta \Psi_2- \frac{4\pi}{3}\left(- Q \mathcal{A}+ 2Q\frac{ f_{,\chi}}{f} - Q_{,\chi} \right) L, \label{W_propApp}\\
 \Psi_2 &=&  \left(\frac{1}{3}\mathcal{A}^2 + \frac{\mathcal{A}f_{,\chi}}{3f} +\frac{1}{3} \mathcal{A}_{,\chi} \right)L^2, \label{zeroShApp}\\
\frac{\kappa}{2} Q &=& \frac{1}{3} \Theta_{,\chi},\\
(\Psi_2+ \frac{4\pi}{3} \rho)_{,\chi}&=&  \frac{4\pi}{3} \Theta Q -3\Psi_2 \frac{ f_{,\chi}}{f}, \label{W_constrApp} \\
\frac{1}{9}\Theta^2 &=& \frac{8\pi}{3} \rho + \frac{1}{6} \lidx{3}{R}.
\end{eqnarray}
where the dot denotes $u^a\nabla_a = (L/N)\partial/\partial t=(L/aN)\partial/\partial \eta$.


\end{appendices}


\bibliography{references.bib}

\end{document}